\pdfoutput=1
\documentclass[reprint,amsmath,amssymb,prd,nofootinbib, superscriptaddress]{revtex4-1}

\usepackage{amsmath}
\usepackage{aas_macros}
\usepackage{graphicx}
\usepackage{dcolumn}
\usepackage{bm}
\usepackage{soul,xcolor}
\usepackage[normalem]{ulem}

\usepackage{xcolor}

\definecolor{colorLink}{rgb}{0.9,0,0} 
\definecolor{colorCite}{rgb}{0,0.7,0} 
\definecolor{colorURL} {rgb}{0,0,0.8} 
\usepackage[colorlinks=true,linktocpage=true,linkcolor=colorLink,citecolor=colorCite,urlcolor=colorURL]{hyperref}
\newcommand{\be}{\begin{equation}}
\newcommand{\ee}{\end{equation}}

\usepackage{colortbl}
\usepackage{fontawesome}
\usepackage{bbold}

\setstcolor{red}

\newcommand{\sk}[1]{}
\newcommand\msun{M_{\odot}}

\newcommand{\h}{\mathbb{h}}

\defcitealias{ias_template_bank_PSD_roulet2019}{R19}

\graphicspath{{./}{figs/}}

\begin{document}


\title{New approach to template banks of gravitational waves with higher harmonics: Reducing matched-filtering cost by over an order of magnitude}

 \author{Digvijay Wadekar}
  \email{jayw@ias.edu}
 \affiliation{\mbox{School of Natural Sciences, Institute for Advanced Study, 1 Einstein Drive, Princeton, NJ 08540, USA}}
 \author{Tejaswi Venumadhav}
\affiliation{\mbox{Department of Physics, University of California at Santa Barbara, Santa Barbara, CA 93106, USA}}
\affiliation{\mbox{International Centre for Theoretical Sciences, Tata Institute of Fundamental Research, Bangalore 560089, India}}
\author{Ajit Kumar Mehta}
\affiliation{\mbox{Department of Physics, University of California at Santa Barbara, Santa Barbara, CA 93106, USA}}
 \author{Javier Roulet}
\affiliation{TAPIR, Walter Burke Institute for Theoretical Physics, California Institute of Technology, Pasadena, CA 91125, USA}
\author{Seth Olsen}
\affiliation{\mbox{Department of Physics, Princeton University, Princeton, NJ 08540, USA}}
\author{Jonathan Mushkin}
\affiliation{\mbox{Department of Particle Physics \& Astrophysics, Weizmann Institute of Science, Rehovot 76100, Israel}}
\author{Barak Zackay}
\affiliation{\mbox{Department of Particle Physics \& Astrophysics, Weizmann Institute of Science, Rehovot 76100, Israel}}
 \author{Matias Zaldarriaga}
\affiliation{\mbox{School of Natural Sciences, Institute for Advanced Study, 1 Einstein Drive, Princeton, NJ 08540, USA}}
 \date{\today}

\begin{abstract}

Searches for gravitational wave events use models, or templates, for the signals of interest. The templates used in current searches in the LIGO--Virgo--Kagra data model the dominant quadrupole mode $(\ell,|m|)=(2,2)$ of the signals, and omit sub-dominant higher-order modes (HM) such as $(\ell,|m|)=(3,3)$, $(4,4)$, which are predicted by general relativity. This omission reduces search sensitivity to black hole mergers in interesting parts of parameter space, such as systems with high masses and asymmetric mass-ratios.
We develop a new strategy to include HM in template banks: instead of making templates containing a combination of different modes, we separately store normalized templates corresponding to $(2,2)$, $(3,3)$ and $(4,4)$ modes. To model aligned-spin $(3,3)$, $(4,4)$ waveforms corresponding to a given $(2,2)$ waveform, we use a combination of post-Newtonian formulae and machine learning tools. In the matched filtering stage,
one can filter each mode separately with the data and collect the timeseries of signal-to-noise ratios (SNR).  This leads to a HM template bank whose matched-filtering cost is just $\approx 3\times$ that of a quadrupole-only search (as opposed to $\approx\! 100 \times$ in previously proposed HM search methods). 
 Our method is effectual and generally applicable for template banks constructed with either stochastic or geometric placement techniques.
 New GW candidate events that we detect using our HM banks and details for combining the different SNR mode timeseries are presented in accompanying papers: \citet{Wad23_HM_Events} and \cite{Wad23_Pipeline} respectively.
 Additionally, we discuss non-linear compression of $(2,2)$-only geometric-placement template banks using machine learning algorithms. 
 \href{https://github.com/JayWadekar/gwIAS-HM}{\faGithub}


 
\end{abstract}
\maketitle

\section{Introduction}\label{sec:intro}
\begin{figure*}
\centering
\includegraphics[width=\textwidth,keepaspectratio=true]{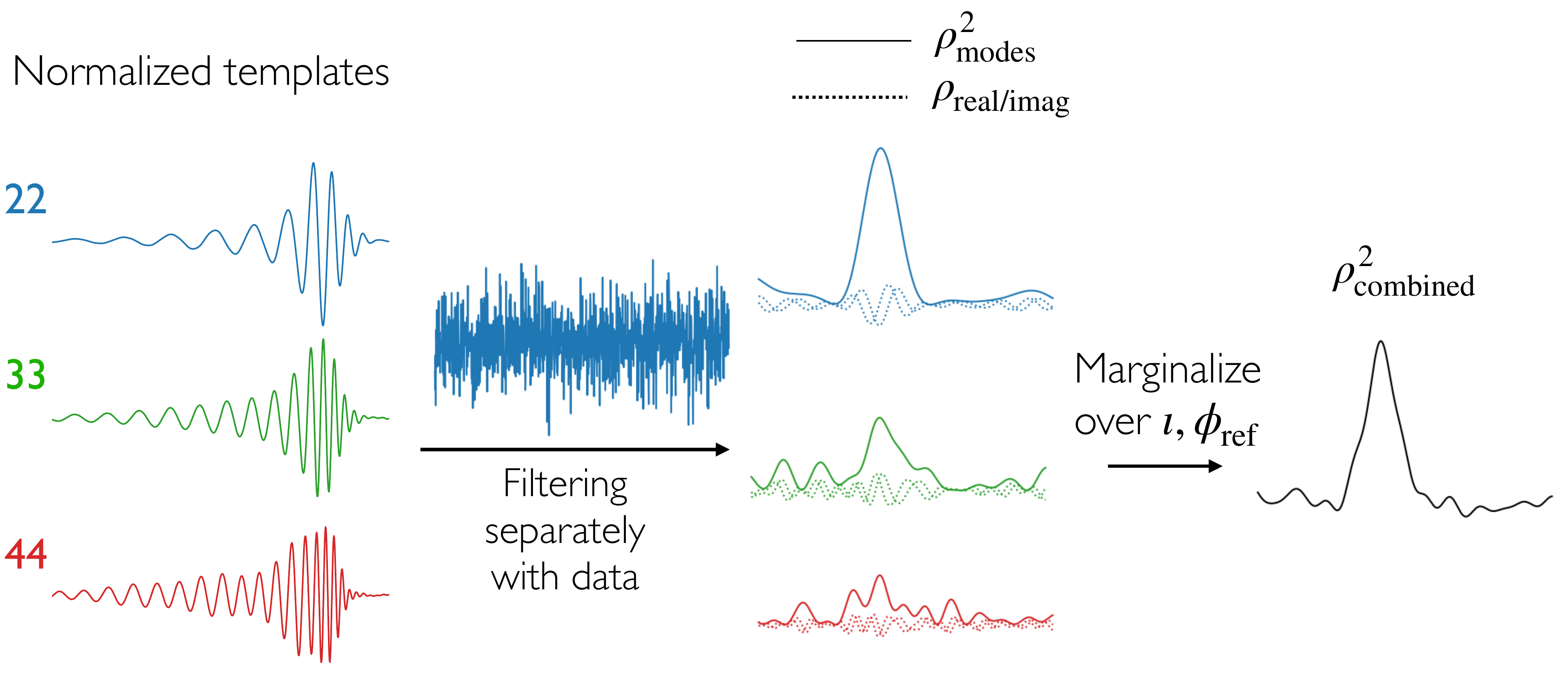}
\caption{
In our template banks, for each set of intrinsic parameters (i.e., masses and spins [$m_1,m_2, s_{1z}, s_{2z}$]), we generate and store the normalized (2,2), (3,3) and (4,4) mode waveforms separately. We then filter the data with the individual modes, which results in a cost increase of just $3\times$ compared to the (2,2)-only case. Note that the $3\times$ factor is significantly smaller than a factor of $\sim 100\times$ occurring when the templates are made with combinations of different harmonics (in the combined case, 
 templates need to span the larger 6D space: [$m_1,m_2, s_{1z}, s_{2z}, \iota, \phi_0$], where $\iota$ is the inclination and $\phi_0$ is the initial reference phase of the binary) \cite{Cha22, Sch23_NF_TemplateBank, Har18}).
We store the output SNR timeseries of each mode and later combine them by marginalizing over $\iota$ and $\phi_\mathrm{ref}$ (details of the marginalization algorithm are given in our companion paper: Ref.~\cite{Wad23_Pipeline}, and we also provide a brief overview in Section~\ref{sec:matched_filtering}).}
\label{fig:Triggering_modes}
\end{figure*}

The bulk of the gravitational wave (GW) detections until now have come from 
template-based searches, which filter strain data from detectors with known waveforms  \cite{O1catalog_LVC2016, gwtc1_o2catalog_LVC2018, lvc_o3a_gwtc2_catalog_2021, gstlal, PYCBCPipeline, mbta_o3a_pastro_andres2022, lvc_o3a_deep_gwtc2_1_update_2021, lvc_gwtc3_o3_ab_catalog_2021, GW230529, ias_o2_pipeline_new_events_prd2020, Wad23_HM_Events, Ols22_ias_o3a, Meh23_ias_o3b, Chi23, NitzCatalog_2-OGC_o2_2020, nitz_o3a_3ogc_catalog_2021, nitz_4ogc_o3_ab_catalog_2021}. These searches rely on having banks of templates with accurate GW waveforms that cover the parameter space within which one intends to search for compact object mergers.

Gravitational wave signals can be decomposed into basis harmonics, or modes, each of which has a characteristic spherical-harmonic angular dependence on the detectors' position on the source's sky \cite{Tho80,Var14}. The quadrupole mode $(\ell,|m|)=(2,2)$ is the leading-order term in the post-Newtonian (PN) multipole expansion, but higher-order modes (HM) such as $(\ell,|m|)=(3,3)$ or $(4,4)$ can be excited to detectable levels close to the merger for binary sources with large and asymmetric masses. Almost all of the currently used template banks include waveforms with only the dominant quadrupole mode \sk{$(\ell,|m|)=(2,2)$ }instead of the full gravitational wave waveform predicted by general relativity \cite{gstlal, PYCBCPipeline,Ols22_ias_o3a,nitz_4ogc_o3_ab_catalog_2021, mbta_o3a_pastro_andres2022,Sak23_TemplateBank_GSTLAL}. 
\sk{While the quadrupole mode is the leading-order term in the post-Newtonian (PN) multipole expansion, higher-order modes (HM) such as $(\ell,|m|)=(3,3)$ or $(4,4)$ are excited to detectable levels for binaries with large and asymmetric masses.}
In addition to ignoring HM, current template banks also assume that the constituent spins 
are aligned with the orbital angular momentum and the binary orbit is nearly circular (i.e., they ignore the effects of spin-orbit precession and orbital eccentricity).

While the effects of HM and precession are ignored in GW searches, they are now being included in most studies estimating source parameters of high-significance triggers obtained from the searches. One justification for this can be that the purpose of searches is merely to identify potentially interesting small segments from the full data stream; one can be more precise in the parameter estimation stage because additional effects like HM and precession are not prohibitively expensive when restricting to a targeted analysis of small data segments.
However, ignoring these additional effects can change which triggers enter the final detection catalog, potentially missing interesting systems when the signal-to-noise ratio (SNR) of these candidates is not very high. Furthermore, the significance estimate of triggers corresponding to a particular class of physical systems (e.g., high mass systems) can be affected (i.e., the false alarm probability of such triggers can increase).

In this paper, we will focus on making banks that include the effects of HM in their templates.
HM can contribute significantly to SNR in the detectors' sensitive band for sources with particular parameters, such as edge-on orientations, unequal mass ratios, and high total mass. 
Hence, searches that only model the quadrupole mode can lose sensitivity to such events \cite{HMeffect_ParameterSpaceDependency_PekowskyPRD2013, HMeffect_RelativeModeSignificance_HealyPRD2013,  Cap14, HMeffect_AlignedSearchImpactCalderonBustilloPRD2016, HMandPrecessionEffect_HeavySearchImpact_CalderonBustilloPRD2017, HMeffect_IMBHsearchImpact_CalderonBustilloPRD2018, Mil21, Har18, Cha22, Sha22,Zha23} (see Figs.~\ref{fig:SNR_Mtot_q} and~\ref{fig:HM_PSD} below). One caveat here is that the sensitive volume of 
asymmetric-mass systems (i.e., mass ratio $q = m_2 / m_1 < 1$) is much smaller than the corresponding equal-mass ($q=1$) systems. However, detecting even a few systems with $q\lesssim 1/5$ can help improve our understanding of some of the compact object formation channels \cite{GW190814, GW190412}. This has motivated previous studies (e.g., \cite{Har18, Cha22, Sch23_NF_TemplateBank}) to construct template banks that include HM (it is worth mentioning that there have also been efforts to include HM in searches without using pre-computed templates but instead using machine learning tools directly \cite{Sha22, Sch22, Tin23}).

One of the biggest hurdles in creating template banks that include the effects of HM (or precession or eccentricity) is that the extra degrees of freedom significantly increase the number of templates. When making aligned-spin template banks with just the $(2,2)$-mode, one only needs to sample over the masses and spins of the binary. However, constructing HM banks requires sampling over two additional parameters: inclination and initial phase ($\iota, \phi_0$), defined at some reference epoch. These parameters modulate the amplitude and phase of HM waveforms relative to the $(2,2)$. Because of these additional degrees of freedom, the number of templates increases by $\sim100\times$ compared to the $(2,2)$-only case \cite{Cha22, Sch23_NF_TemplateBank, Har18} (with the matched-filtering cost of the search also increasing by the same factor).

In this paper, we develop a different strategy for constructing HM template banks. For each set of binary masses and spins, we store the (2,2), (3,3), (4,4) mode waveforms separately in our bank.
We also matched-filter the data with these modes separately and store the three SNR timeseries as shown in Fig.~\ref{fig:Triggering_modes}. We can then combine the three SNR timeseries to marginalize over the angles $\iota, \phi_0$ and get a statistic which can be used to collect and rank triggers. As the matched-filtering step involves three computations for every one computation in the analogous $(2,2)$-only case, the cost of the filtering step is increased by a factor of three. Note that we have an additional step in our case: combining the different SNR timeseries by maximization (or marginalization) over $\iota$ and $\phi_0$. However, the cost of this additional step is much smaller than the matched-filtering step. We discuss combination of the SNR timeseries in detail in our first companion paper \cite{Wad23_Pipeline}, where we also introduce new detection statistics to mitigate the increase in the number of background triggers (which is a result of the extra degrees of freedom introduced by HM
in the template bank \cite{Cap14}). 
Our second companion paper \cite{Wad23_HM_Events} details the new binary black hole (BBH) mergers found using our search methodology in the LIGO--Virgo O3 data.


\begin{figure*}
\centering
\includegraphics[width=0.335\textwidth,keepaspectratio=true]{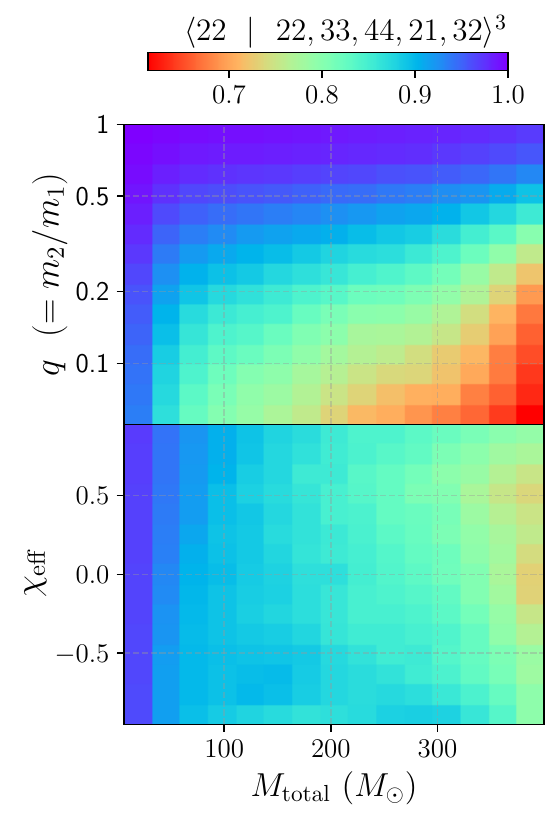}
\includegraphics[width=0.31\textwidth,keepaspectratio=true]{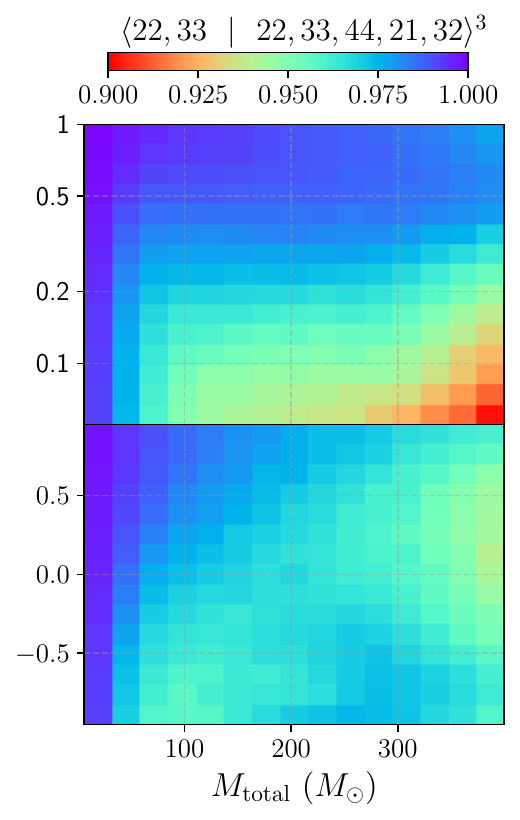}
\includegraphics[width=0.31\textwidth,keepaspectratio=true]{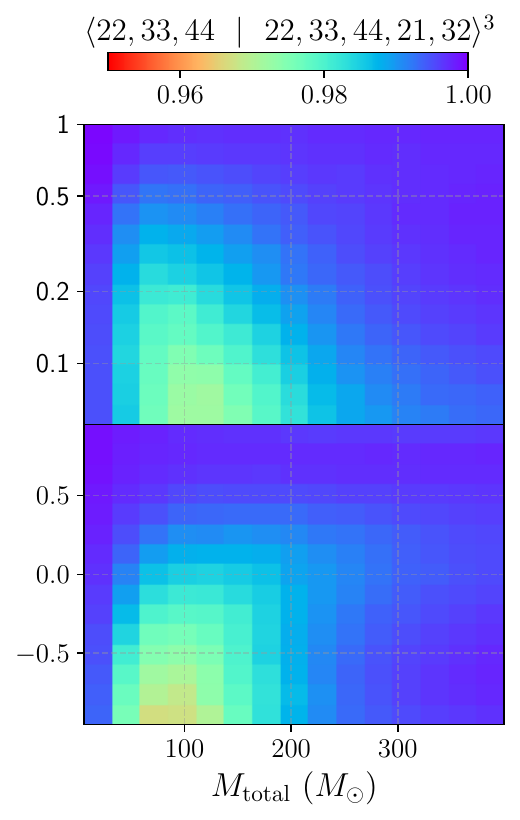}
\caption{Ignoring HM in the waveform templates can cause a loss of waveform overlap with the true signal, which in turn leads to a reduction in the detection volume of systems. In this figure, we show the fractional detection volume of systems (calculated by the cube of the waveform overlap in a simplistic scenario) in particular regions of parameter space. In the \textit{left} panel, we include only the dominant $(2,2)$ mode in the templates and see that the detection volume loss is larger for high-mass and asymmetric mass ratio systems (see Fig.~\ref{fig:HM_PSD} for the reasoning behind the high mass behavior). The {\it center} and {\it right} panels show the behavior when additional $\ell = |m|$ modes are included in addition to $(2,2)$ (omitting the $\ell\neq |m|$ modes: $(2,1)$, $(3,2)$).
We find that the fractional loss in volume is $\lesssim 4$\% in the given parameter space, showing that adding only the (3,3) and (4,4) modes to our template banks is roughly sufficient.
In all the panels, we have averaged over inclination accounting for the brightness of the 22 waveform.
Note that the range of the color bar is significantly different in the three panels.}
\label{fig:SNR_Mtot_q}
\end{figure*}

\begin{figure}
\includegraphics[scale=0.7,keepaspectratio=true]{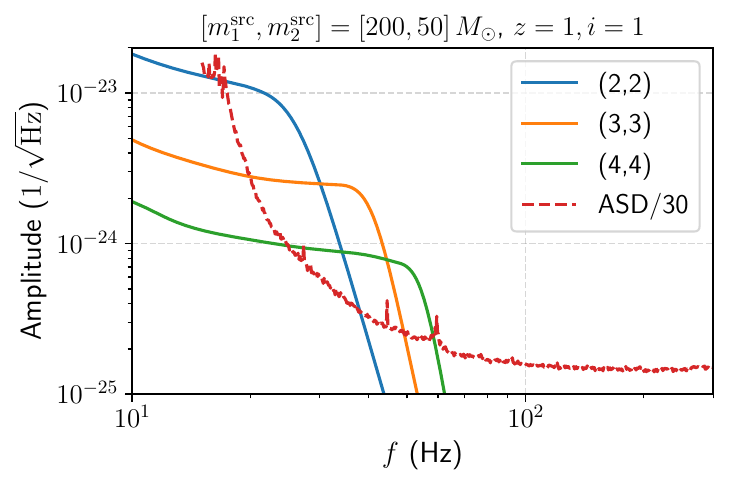}
\caption{As seen in Fig.~\ref{fig:SNR_Mtot_q}, the fractional contribution of HM to SNR increases significantly for high-mass binaries. This is because the noise ASD (amplitude spectral density) increases at low-frequencies, which leads to preferential downweighting of the (2,2) mode SNR, especially for high masses where the (2,2) signal is dominated by lower frequencies (while the HM, being at higher frequencies, are less affected). The reference noise ASD shown here corresponds to the O3 run and has been used throughout this paper. The parameters of the system used to generate the mode curves are shown in the title of the plot.}
\label{fig:HM_PSD}
\end{figure}

We discuss the formalism and simulations of waveforms containing HM in Section~\ref{sec:GW_HM}. In Section~\ref{sec:22}, we describe our model for making $(2,2)$-only banks. In Section~\ref{sec:hm_model}, we describe our model for making amplitudes and phases of HM templates using post-Newtonian expansion and machine learning techniques. In Section~\ref{sec:matched_filtering}, we 
briefly describe the process of matched-filtering the data with our template bank. 
In Section~\ref{sec:Effectualness}, we show that our HM template bank is effectual. 
In Section~\ref{sec:discussion}, we discuss an application of our technique to stochastic placement based template banks, and we conclude in Section~\ref{sec:conclusions}.

\section{Gravitational wave signals including higher harmonics}
\label{sec:GW_HM}

\subsection{Generating waveform samples for the bank}
\label{sec:wf_samples}
To generate aligned-spin waveforms which include HM, we use the recently developed \texttt{IMRPhenomXHM} model \cite{Gar20}. This approximant calibrates analytic PN-based expressions of different multipoles (which are accurate in the early-inspiral regime of the binary evolution) with numerical-relativity simulations (which more accurately describe the merger-ringdown regime)\footnote{We also attempted generating waveforms with the \texttt{IMRPhenomXPHM} approximant, which includes both HM and precession. However, we found a few inconsistencies in the case when the spins are nearly aligned ($s_{1r}\sim 0.01$) versus exactly aligned ($s_{1r}=0$), which could be due to borderline cases between the separate parameter space regions where the approximant was calibrated. This led to anomalies in our SVD basis for the phases (which are calculated later in Section~\ref{sec:22_phases}).}. In addition to $(2,2)$, we use all the currently available modes in \texttt{IMRPhenomXHM}: $(3,3)$, $(4,4)$, $(2,1)$, $(3,2)$. Hereafter we write $(\ell, m)$ mode as just $\ell m$ for brevity.

We only generate BBH templates in this study and leave systems containing neutron stars for a different study.
In order to model our template banks, we simulate waveforms in the following parameter range:
\be\begin{split}
3 M_\odot < m_2 &< m_1 < 400 M_\odot \\
1/18 &< q < 1\\
|\chi_1|, |\chi_2| &<0.99
\end{split}\label{eq:prior}\ee
where the masses are in detector frame (we do not use separate cuts for redshift).
The motivation for the mass ratio ($q$) cut is that  \texttt{IMRPhenomXHM} approximant was calibrated to numerical relativity (NR) waveforms up to $q>1/18$, while the cut on the individual spins is due to the relativistic Kerr limit.
To construct our template banks, we simulate a sample of $\sim 10^{6}$ waveforms (each corresponding to the physical parameter set $[m_1, m_2, \chi_{1,z}, \chi_{2,z}]$). These samples are broadly distributed across the parameter space in Eq.~\ref{eq:prior}. Similar to other templated searches, we need to assume an astrophysical prior distribution in order to give weights to the templates in our banks. We pick a simple prior: uniform prior in $q$ for $1/18 < q < 1$; flat prior for $\chi_\mathrm{eff}$ bounded within $-0.95<\chi_\mathrm{eff}<0.95$ (with $|\chi_1|, |\chi_2| <0.99$); and a power law distribution in the total redshifted mass: $P(M_\mathrm{tot})\propto M_\mathrm{tot}^{-2}$ (we do not assume a separate prior over redshift).

We first quantify the fractional loss in the recovered SNR if we ignore the HM in our templates. To calculate the overlap of two waveforms, we use
\be
\langle h_i | h_j \rangle = 4 \int_0^\infty {\rm d}f \frac{h^*_i(f)h_j(f)}{S_n(f)}
\label{eq:inner_prod}
\ee
where $S_n(f)$ is the one-sided power spectral density (PSD). Throughout this paper, we use a reference PSD that is obtained by applying Welch's method~\cite{1161901} over 50 random O3 LIGO data files and taking the $10$th percentile of the sample of PSDs in order to downweight the distortions arising from spectral lines and loud glitches.

An important factor in the calculation of loss of SNR is the inclination $\iota$ of the binary. This is because HM become more important when $\iota$ is away from 0 or $\pi$ (zero corresponds to a face-on binary). However, the observable volume of 
systems with inclination closer to edge-on is diminished, since 
the brightness of the 22 mode starts to decrease (typically the 22 mode is brighter than HM so 
it drives our choice for the inclination distribution).
We therefore simulate test waveforms using the following probability distribution, which includes the observable volume effect (see Fig.~4 of \cite{Sch11}):
\be \label{eq:p_iota}
P(\iota) = \left[ \frac{1}{8}(1+6 \cos^2 \iota +\cos^4 \iota)\right]^{3/2}\, \sin \iota.
\ee
Note that similar comparisons have already been done in previous studies \cite{HMeffect_ParameterSpaceDependency_PekowskyPRD2013, HMeffect_RelativeModeSignificance_HealyPRD2013,  Cap14, HMeffect_AlignedSearchImpactCalderonBustilloPRD2016, HMandPrecessionEffect_HeavySearchImpact_CalderonBustilloPRD2017, HMeffect_IMBHsearchImpact_CalderonBustilloPRD2018, Mil21, Har18, Cha22,Sin23_HM_Populations} for fixed $\iota$, whereas here we show the case when the $\iota$ dependence has been marginalized over. We show the results of the fractional loss in SNR in the left panel of Fig.~\ref{fig:SNR_Mtot_q} (the color shows the overlap given by $\big|\langle h_{22}|h\rangle \big|/ \sqrt{\langle h_{22}|h_{22}\rangle\langle h| h\rangle}$ where $h$ includes all modes: 22, 33, 44, 21, 32). Note that the detection volume shown in the figure does not include the effect of loss in sensitivity from increasing background triggers due to the larger number of waveforms \cite{Cap14}. To thoroughly study this issue, we plan to do an injection-based study using the \texttt{IAS-HM} pipeline\footnote{\url{https://github.com/JayWadekar/gwIAS-HM}} and explicitly measuring the effect of adding HM on the effective volume-time ($VT$) of the search.

We also perform another test where we add two of the HM (33 and 44) alongside 22 and test the overlaps with the full waveform.
From the right panel of Fig.~\ref{fig:SNR_Mtot_q}, we see that using these three modes gives fairly accurate overlap with the full waveform throughout the parameter space. 

The model for the gravitational wave strain in the detector is given by $h_\mathrm{det} = F_+ h^+ + F_\times h^\times$ where $F_+, F_\times$ are the detector's response functions, which depend on sky location and polarization angle $\psi$ of the binary \cite{Var14, Mil21}. The waveform polarization modes are given by
\be\begin{split}
h^+(f) -ih^\times (f) = \frac{D_0}{D_L}  \sum_{\ell,m\geq2} {}_{-2}Y_{\ell m} (\iota,\phi_0)\, h_{\ell m} (f)
\end{split}\ee
where $D_L$ is the luminosity distance, $D_0$ is the reference distance at which the \sk{spherical} harmonic modes $h_{\ell m}$ were generated, and ${}_{-2}Y_{\ell m}$ corresponds to the spin-weighted spherical harmonics. Explicitly expanding the expressions for ${}_{-2}Y_{\ell m}$ for the case of 22, 33 and 44, we get \cite{Mil21}
\begin{equation}
\begin{split}\label{eq:predictedwf}
  h_\mathrm{det}(f) =& \frac{D_0}{D_L} \left[ F_{+} \frac{1 + \cos^2\iota}{2} \, - \, i\cos \iota\, F_{\times}\right]\times
   \\
   \Big[& e^{2 i \phi_0}  h_{22}(f)-\sqrt{\frac{21}{10}} \sin \iota\, e^{3 i \phi_0}  h_{33}(f)\\&+ \sqrt{\frac{63}{20}} \sin^2 \iota\, e^{4 i \phi_0} h_{44}(f)\Big].
\end{split}
\end{equation}

One can see that the detector responses conveniently factor out as an overall normalization term \cite{Mil21}. This is however not the case if $\ell \neq |m|$ modes are included in this expression, and we discuss this point further in Appendix~\ref{apx:Polarizations}. It is also evident that, if one only considers the 22 waveform in Eq.~\eqref{eq:predictedwf}, there is a degeneracy between $\iota, \phi_0$ and $\psi$ (included in the detector responses). In that case, one does not need to make different templates to account for these variables. However, when including HM, this degeneracy is broken and therefore, in order to model the full 22+HM waveform, it is necessary  to make different templates for different values of $\iota$ and $\phi_0$. This is indeed the approach taken by previous HM template banks \cite{Har18, Cha22, Sch23_NF_TemplateBank}.
We, however, take a different approach in this paper: we
construct templates for the 22, 33 and 44 modes separately, so our templates do not include any information about $\iota$ and $\phi_0$. Instead, we include the dependence on $\iota$ and $\phi_0$ at a later stage in the pipeline when we combine the SNR timeseries after matched-filtering with the templates (see Ref.~\cite{Wad23_Pipeline}).
We discuss the model for our templates in the next two sections.

\section{Our Quadrupole bank model}
\label{sec:22}
In our method, there is a template for the $(3, 3)$ mode and a template for the $(4, 4)$ mode associated with every $(2, 2)$ template. We first create our (2,2) bank independent of HM and discuss our model in this section (we will discuss modeling HM templates in the next section).

There are two broad classes of (2,2) template banks in the literature. One is stochastic placement, where the bank consists of waveforms which are randomly sampled (see e.g., \cite{PYCBCPipeline, Aji14_TemplateBank}).
We will use the alternative approach of geometric placement, where a regular lattice is constructed based on the metric induced by the inner product of the waveforms, Eq.~\eqref{eq:inner_prod} (see e.g., \cite{Brown12_GeometricPlacement, ias_template_bank_PSD_roulet2019, Han23_GeometricPlacement}). It is worth noting that our methodology of including HM in the template banks is equivalently applicable to 22 banks constructed with stochastic placement (we will come back to this point later in Sec.~\ref{sec:stochastic}). 
To build our 22 banks, we use a method based on 
Ref.~\cite{ias_template_bank_PSD_roulet2019} (hereafter \citetalias{ias_template_bank_PSD_roulet2019}).
\citetalias{ias_template_bank_PSD_roulet2019} shows that decomposing the frequency-domain waveform into an amplitude and an unwrapped phase term is advantageous because the amplitude $A_{22}(f)$ and phase $\Psi_{22}(f)$ separately depend on the parameters of the binary in a smooth way (see Fig.~1 of \citetalias{ias_template_bank_PSD_roulet2019}).

\subsection{Modeling amplitudes and separation into banks}

\begin{figure}
\centering
\includegraphics[width=0.45\textwidth,keepaspectratio=true]{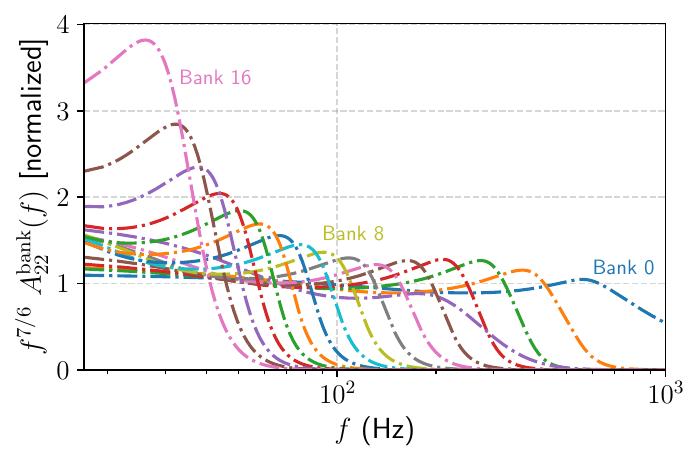}
\includegraphics[width=0.45\textwidth,keepaspectratio=true]{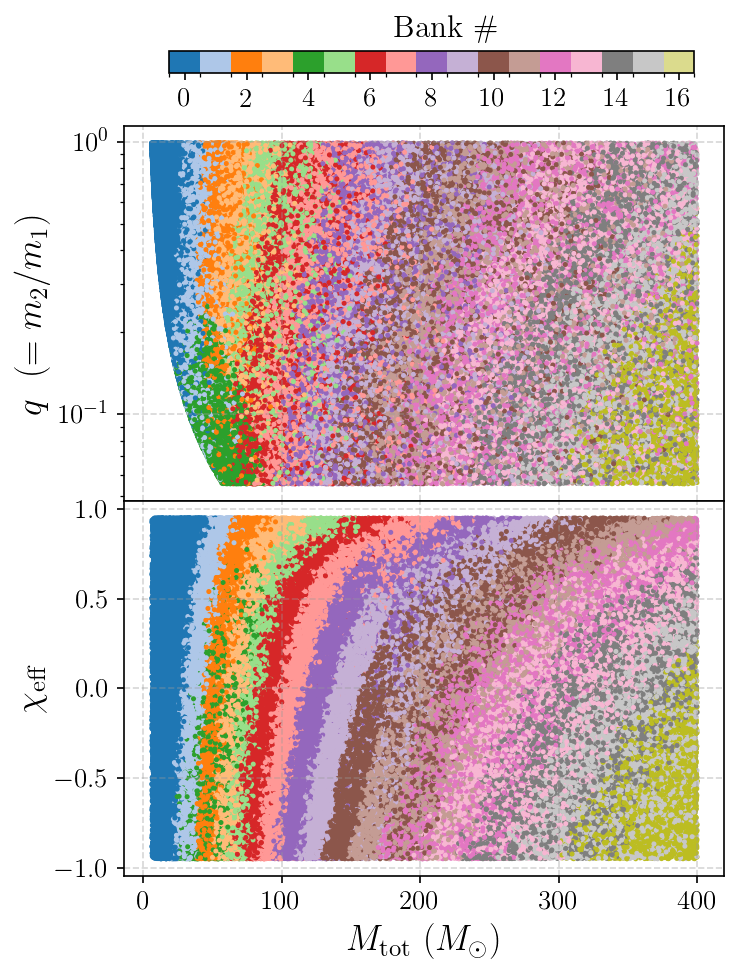}
\caption{The template banks in our analysis are split a according to the normalized waveform amplitudes of the (2,2) mode. 
\textbf{Top:} Normalized amplitude profiles corresponding to our banks, where we see that the different banks are roughly distinguished by the cutoff frequencies of the waveforms.
\textbf{Bottom:} Physical parameters corresponding to the different banks.}
\label{fig:bank_parameters}
\end{figure}

To divide the binary mass and spin samples into banks, we first find the normalized amplitudes (i.e., $\langle 
\overline{A_{22}}| \overline{A_{22}}\rangle=1$) of the waveforms from IMRPhenomXHM corresponding to these samples.
Our banks are made by requiring that waveforms with similar amplitude profiles are grouped together in a single bank. We pass $\overline{A_{22}}(f)$ waveforms over the entire parameter space through a \texttt{KMeans} algorithm~\cite{scikit_learn}. This algorithm identifies centroids in the space of waveform amplitudes and each input waveform amplitude is then associated to a particular centroid. Each centroid corresponds to an amplitude profile which we assign as the reference amplitude of that bank: $A^\mathrm{bank}_{22} (f)$ (note that we also normalize these amplitudes such that $\langle A^\mathrm{bank}_{22}|A^\mathrm{bank}_{22}\rangle=1$). The choice of the centroids is made such that the average overlap between the amplitudes associated to that bank, given by $\langle \overline{A_{22}} (f)| A^\mathrm{bank}_{22} (f)\rangle$, is maximized. We choose the number of banks to be 17 based on the criteria that the overlap is $>0.96$ for more than 99.9\% of the waveforms associated to that bank. We show the profiles of $A^\mathrm{bank}_{22} (f)$ for our banks in the top panel of Fig.~\ref{fig:bank_parameters}. In the bottom panel, we show the parameters associated to each bank based on their waveform amplitude profiles. The curved boundary seen in the bottom left of the $q$--$M_\mathrm{tot}$ plot is because of the condition that the secondary black hole has mass $> 3 \,\msun$.

We further choose to divide low-mass banks into subbanks based on $\mathcal{M}_\mathrm{chirp}$. This is because waveforms with different $\mathcal{M}_\mathrm{chirp}$ have different duration. Non-Gaussian noise is known to affect short-duration waveforms much more than long-duration waveforms, and hence this division helps in isolating long-duration waveforms from the additional background. We choose the number of subbanks within each bank such that the duration of waveforms in each subbank is roughly similar to within a factor of two. We show the $\mathcal{M}_\mathrm{chirp}$ cuts for each subbank in Table~\ref{tab:Ntemplates}.
We also show the number of templates in each bank. 
Note that smaller the binary mass, the larger the number of orbital cycles seen by current detectors, and therefore more templates are needed to model that region ($N_\mathrm{templates}\propto \mathcal{M}_\mathrm{min}^{-8/3}$~\cite{Owen:1998dk}). If we assume the same SNR threshold across all the banks to collect the matched-filtering triggers, we will run into the problem that there will be vastly larger number of background triggers in low-mass banks compared to high-mass banks due to a difference in the number of templates. To get around this issue, we employ different SNR collection thresholds in different banks (e.g., we use $\rho^2_\mathrm{collection}=26$ for BBH-0, whereas for BBH-8 and above, we use $\rho^2_\mathrm{collection}=20$).

Note that the banks in \citetalias{ias_template_bank_PSD_roulet2019} were divided using $\mathcal{M}_\mathrm{chirp}$ values in the entire parameter range (and subbanks were defined based on the shapes of the normalized amplitudes of waveforms).
For low-mass binaries, the waveform is inspiral-dominated and thus the division by $\mathcal{M}_\mathrm{chirp}$ is appropriate; however for high-mass binaries, the waveform is dominated by the merger and the ringdown phase, hence the waveform behavior depends primarily on $M_\mathrm{tot}$ instead of $\mathcal{M}_\mathrm{chirp}$. Therefore, in our case, our bank division is only based on the normalized amplitudes in the high-mass case, and we forego further division by $\mathcal{M}_\mathrm{chirp}$. For the low mass case, our subbanks are roughly similar to the subbanks of \citetalias{ias_template_bank_PSD_roulet2019}.



\subsection{Modeling phases}
\label{sec:22_phases}
We model the phases of waveforms based on the singular value decomposition (SVD) method introduced in \citetalias{ias_template_bank_PSD_roulet2019}. For \texttt{IMRPhenomXHM} waveforms associated with each bank, we first extract their unwrapped phases, subtract the average phase and also orthogonalize the phases with respect to time offsets.
We perform an SVD of the phase profiles to obtain an orthonormal basis for the phase functions. We find that the phase profile can be well modeled by a low dimensional space:
\be
\Psi^\mathrm{model}_{22} (f) = \langle\Psi_{22}\rangle_\mathrm{subbank} (f) + \sum_{i=0}^\mathrm{few} c^{(22)}_i \Psi^\mathrm{SVD}_i (f),
\label{eq:phases_22}
\ee
where $\Psi_i$ corresponds to the orthonormal basis functions from the SVD, and $\langle \Psi \rangle_\mathrm{subbank}$ denotes the average over physical waveforms in the particular subbank.
It is worth noting that, for computing SVDs of the phases, we give a weight to each frequency-bin proportional on the fractional SNR in the bin: $w_k\propto A^\mathrm{bank}_{22}(f_k) \sqrt{\Delta f_k/ S_n(f_k)}$, where $k$ corresponds to the frequency bins and $\Delta f_k$ is the bin width (see \citetalias{ias_template_bank_PSD_roulet2019} for further details).
We also choose a different lower and upper frequency cutoff for different banks. These cutoffs are chosen such that the resulting loss in SNR$^2$ ($1-4\int_{f_\mathrm{min}}^{f_\mathrm{max}} {\mathrm d}f A^\mathrm{bank}_{22}(f)/\sqrt{S_n(f)}$) is lower than $\sim$1\%. This gives $f_\mathrm{min}$ between 19--24\,Hz for different banks, but $f_\mathrm{max}$ varies significantly among banks as they have different cutoff frequencies (see the top panel of Fig.~\ref{fig:bank_parameters}).

\begin{figure}
\centering
\includegraphics[width=0.45\textwidth,keepaspectratio=true]{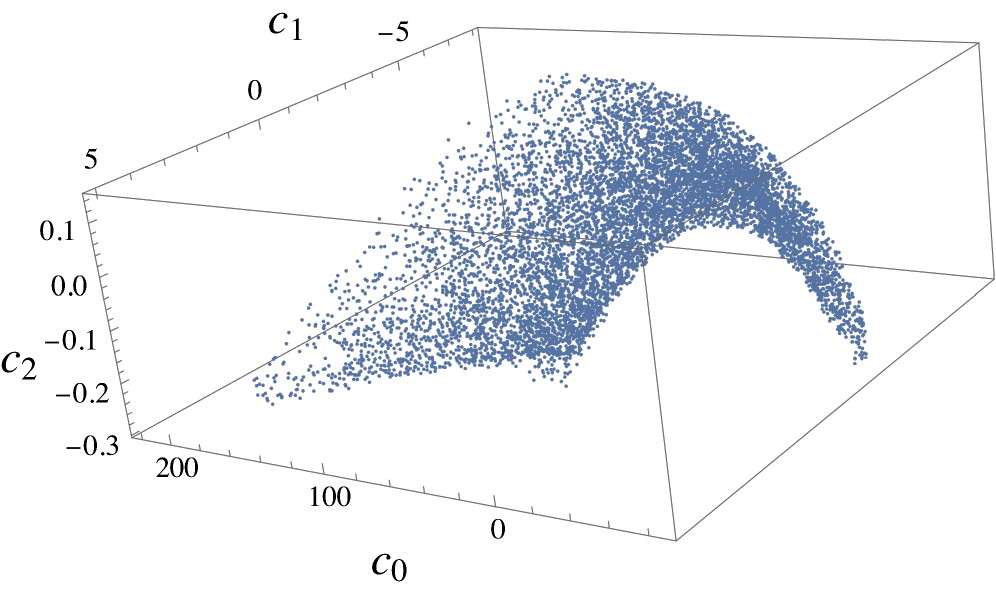}
\includegraphics[width=0.48\textwidth,keepaspectratio=true]{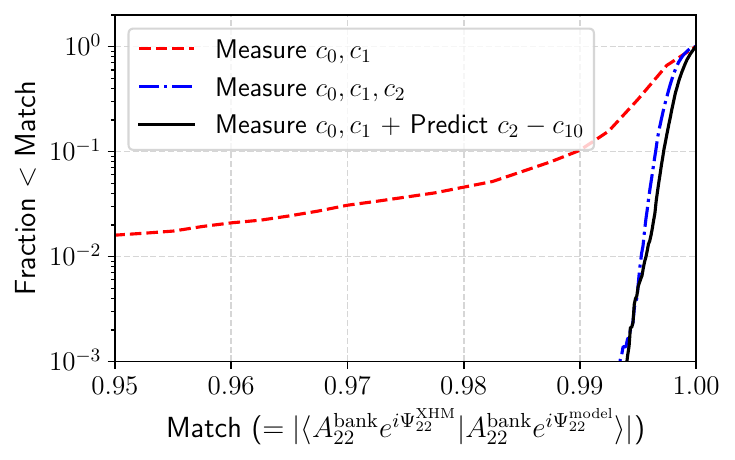}
\caption{We model the phases of our 22 templates using a low-dimensional basis constructed using SVD (see Eq.~\eqref{eq:phases_22}).
The dimensionality of our banks is set by the number of SVD coefficients $c_i$ needed to get effectualness above a particular threshold. \textbf{Top:}  The top three SVD coefficients corresponding to physical waveforms follow a curved hypersurface (showing that $c_0, c_1, c_2$ are not independent, but there is a redundancy which leads to a spurious increase in number of dimensions of our subbanks). \textbf{Bottom:} Histograms of waveform matches for different case of $c_i$ dimensions. Upon modeling $c_{i>2}$ as a function of $c_0, c_1$ using a machine learning tool called random forest regressor (RF), we find that all our quadrupole subbanks can be compressed to two dimensions. Overall, using the RF enables us to search with $\sim 40$\% times fewer templates while keeping a similar level of effectualness.}
\label{fig:RF_calpha}
\end{figure}

For low-mass waveforms, $c^{(22)}_0$ is found to be approximately proportional to $\mathcal{M}_\mathrm{chirp}^{5/3}$, while $c^{(22)}_1$ represents a combination of $q$ and $\chi_\mathrm{eff}$ (corresponding to the well-known degeneracy in the 22 waveform).
The advantage of the model in Eq.~\eqref{eq:phases_22} is that, for nearby templates, the $c_i$ space serves as a Euclidean metric for the mismatch between templates in a particular subbank [\citetalias{ias_template_bank_PSD_roulet2019}]:

\be
\langle h(\bm{c})|h(\bm{c}+\bm{\delta c})\rangle \simeq 1 - \frac{1}{2} \sum_i \delta c^2_i
\label{eq:mismatch_calpha}\ee
 We therefore define a grid in the $\bm{c}$ space and each grid point corresponds to a 22 template in our search. The number of dimensions to include in our grid is chosen based on calculating the match between the model in Eq.~\eqref{eq:phases_22} with the phases of \texttt{IMRPhenomXHM} waveforms in a particular subbank given by
$|\langle A^\mathrm{bank}_{22}(f) e^{i\Psi_{22}(f)}| A^\mathrm{bank}_{22}(f) e^{i\Psi^\mathrm{model}_{22}(f)} \rangle|$.
The histogram of the matches corresponding to the BBH-0,0 subbank (which has the largest number of templates among all subbanks) is shown in Fig.~\ref{fig:RF_calpha}. We can see that two $c_i$ dimensions are not sufficient, but three are enough to have the match $>0.97$ for over 99\% of the waveforms.

It is however important to note that SVD is a linear dimensional reduction tool that we used to model the hypersurface of the physical 22 waveform phases. Using SVD is optimal if the physical hypersurface is linear, but we find this is not true in our case, as is evident from the top panel of Fig.~\ref{fig:RF_calpha}. This can lead to over-predicting the number of dimensions in our waveform model, which would introduce inefficiencies in our search. Note that linear dimensionality reduction tools like SVD or PCA are used extensively in numerous areas of astrophysics, e.g., \cite{Phi21_PCA, Cab02_PCA, Lia23_EHT}; it is important to be aware that the linearity assumption used in these techniques can cause them to be sub-optimal.

A few possible solutions to overcome our problem are to use the kernel trick or use non-linear deep neural network models like autoencoders. However, we want to preserve the Euclidean $c_i$ metric for waveform mismatches, so we use the same SVD basis as earlier but add a machine learning tool called random forest regressor (RF) to reduce the redundancies in the $c_i$ space. For RF, we use the publicly available package \texttt{Scikit-Learn}\footnote{Random forest: \url{https://scikit-learn.org/stable/modules/generated/sklearn.ensemble.RandomForestRegressor.html}} and train the model to predict the higher-order SVD coefficients as a function of the two lowest-order ones:
\be
\{c^{(22)}_2, \ldots , c^{(22)}_{10} \} = \mathrm{RF} (c_0^{(22)}, c_1^{(22)})
\ee
The result for the case of BBH-0,0 is shown in the black line in the bottom panel of Fig.~\ref{fig:RF_calpha}.
This helps in cutting down the dimensionality of the $c_i$ space to two dimensions for all banks and subbanks in our case (in contrast to \citetalias{ias_template_bank_PSD_roulet2019}, where a few of the banks had 3$-$4 dimensions). Furthermore, comparing our template bank size with and without including the RF, we find that RF helps in reducing the number of templates by a factor of $\sim 40$\% while keeping a similar level of effectualness (after including the effect of waveform optimization as described in section~III H of \cite{ias_pipeline_o1_catalog_new_search_prd2019}). We leave exploration of alternative dimensionality reduction techniques like autoencoders to a future study. Finally, along the $c_0^{(22)}$ and $c_1^{(22)}$ dimensions, not all the points of the rectangular grid may describe physically viable waveforms. We therefore use the method adopted in \citetalias{ias_template_bank_PSD_roulet2019} of rejecting the grid points which do not have a physical waveform in their vicinity (for reference, see Fig.~4 of \citetalias{ias_template_bank_PSD_roulet2019}).

\begin{table*}
\centering
\begin{tabular}{p{1.5cm}p{4cm}p{4cm}}
\hline
Bank & N$_\mathrm{templates}$ in subbanks& $\mathcal{M}_\mathrm{bin-edges}$ of subbanks \\ \hline
BBH-$0$  & 10442, 1702, 774, 465, 280 & 2.6, 5.3, 6.8,  8.4, 10.7, 19.8\\
\rowcolor{gray!10}BBH-$1$ & 1902, 318, 157 &  5.8, 11.9, 15.8, 27.3\\
BBH-$2$ &755, 94, 55 & 8.9, 18.5, 23.,  38.6\\
\rowcolor{gray!10}BBH-$3$ & 263, 53, 44& 12.,  22.4, 27.4, 49.6\\
BBH-$4$ &1782, 422, 225 & 5.8, 10.8, 13.5, 19.5\\
\rowcolor{gray!10}BBH-$5$ & 174, 67, 63 & 14.7, 23.9, 30.3, 64.2\\
BBH-$6$ & 249, 115, 76 & 10.3, 21.,  29.5, 75.9\\
\rowcolor{gray!10}BBH-$7$ & 145, 74 & 12.9, 30.6, 92.4\\
BBH-$8$ & 73, 31 & 15.1  39.9, 108.3\\
\rowcolor{gray!10}BBH-$9$ & 37 & 18.5, 127.3\\
BBH-$10$ & 19 & 21.1, 149.1\\
\rowcolor{gray!10}BBH-$11$ & 24 & 24.7, 168.6\\
BBH-$12$ & 5 & 28.4, 173.5\\
\rowcolor{gray!10}BBH-$13$ & 6 & 32.6, 173.8\\
BBH-$14$ & 3 & 37.8, 173.6\\
\rowcolor{gray!10}BBH-$15$ & 3 & 43.2, 173.7\\
BBH-$16$ & 3 & 51.8, 166.\\ \hline
Total & 20890 & \\
\hline
\end{tabular}
\caption{Details of our 22 template banks. The banks are split according to 22 amplitudes (see Fig.~\ref{fig:bank_parameters}) and the subbanks are split according to the chirp mass ($\mathcal{M}$). The lower and upper bin-edges in $\mathcal{M}$ are given in the right column. From BBH-9 onwards, we only have a single subbank per bank.}
\label{tab:Ntemplates}
\end{table*}


It is worth mentioning that the quadrupole bank creation method outlined here has been used recently to also create template banks for an inspiral-only search targeting exotic compact objects with large tidal deformabilities \cite{Chi23}, and also for a search targeting intermediate mass-ratio BBH mergers \cite{Che24_IMRI_search}.

\section{Modeling 33 and 44 waveforms using PN expansion and machine learning}
\label{sec:hm_model}

In this section, we derive effective models for higher harmonics.
The Fourier domain waveforms for different harmonics can be calculated under the stationary phase approximation (SPA) as (see Eq.~(11) of \cite{Mis16}):
\be
h_{\ell m}(f)=\frac{M^2\pi}{D_0} \sqrt{\frac{2 \eta}{3}} V_m^{-7/2} e^{i(m\Psi_\mathrm{SPA}(V_m)+\pi/4)} H_{\ell m}
\label{eq:SPA}
\ee
where $V_k\equiv (2\pi M f/k)^{1/3}$ is the PN parameter in Fourier space (corresponding to the velocity of the binary) and $\eta \equiv m_1m_2/M^2$ is the symmetric mass ratio.

\subsection{Model for amplitudes}

\begin{figure}
\centering
\includegraphics[width=0.48\textwidth,keepaspectratio=true]{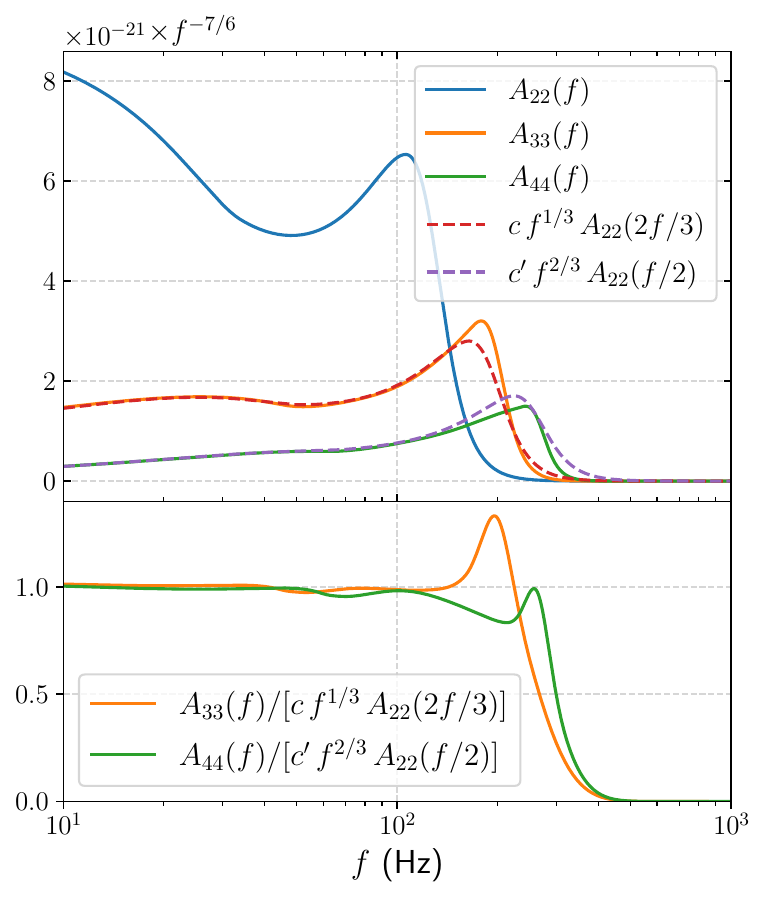}
\caption{Top panel shows amplitude profiles of different harmonics for an arbitrary waveform in bank BBH-8,0 with parameters: $[m_1,m_2,\chi_1,\chi_2] = [87.5, 20.1, -0.46, -0.7]$. Approximations to HM amplitudes derived from low-order PN expansion (see Eq.~\eqref{eq:AmpFormulae}) are shown in dashed. Interestingly, we see that the approximate formulae are fairly accurate even in the quasi-linear regime close to the merger. We show the ratios of the formulae output to the actual waveforms in the bottom panels. Our waveform model for the HM amplitudes is motivated by these formulae and given in Eq.~\eqref{eq:HM_amplitudes}.}
\label{fig:AmpModel}
\end{figure}

\begin{figure}
\centering
\includegraphics[width=0.48\textwidth,keepaspectratio=true]{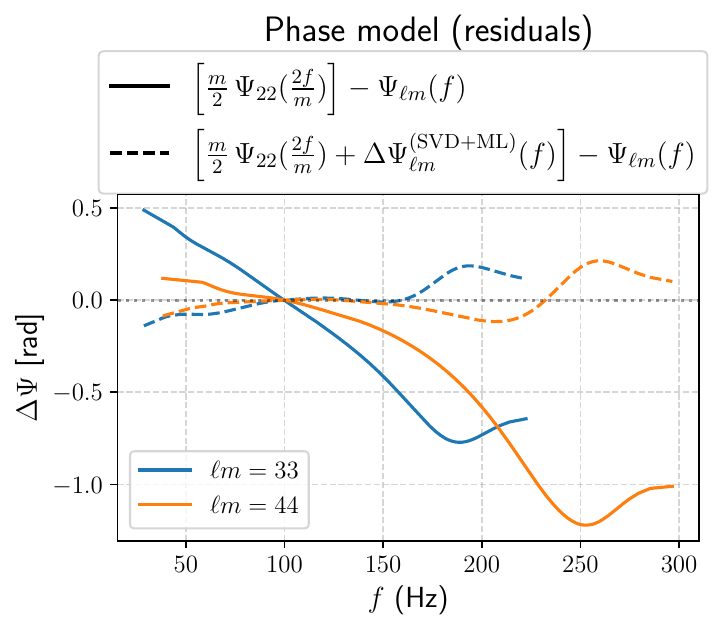}
\caption{Similar to Fig.~\ref{fig:AmpModel}, but showing the accuracy of our model of phases of 33 and 44 waveforms for the same binary system. In the solid lines, we show a simplistic model motivated by low-order PN expansion. The phase residuals are not sufficient for our analysis, therefore we explicitly model the phase residuals from the simple relation using machine learning (ML). To set up our ML model, we simulate similar phase residuals for arbitary waveforms from \texttt{IMRPhenomXHM}, compress the residuals to a few SVD components, and then model the SVD coefficients using a random forest regressor (see Eqs.~\ref{eq:HM_phase_residuals}$-$\ref{eq:HM_calpha}). Applying our model to this particular binary system, we obtain the comparatively more accurate results shown in dashed.}
\label{fig:PhaseModel}
\end{figure}

The various PN-based amplitude terms ($H_{\ell m}$) are given in Appendix~E of \cite{Gar20} (which is an updated version of section~3 of Ref.~\cite{Mis16}). The lowest-order terms for the three multipoles we are interested in are:
\be\begin{split}\label{eq:Hlm}
H_{22}\equiv& -1+ \left( \frac{323}{224}-\frac{451\eta}{168} \right) V^2_2+\mathcal{O}(V^3_2)\\
H_{33}\equiv& -\frac{3}{4}\sqrt{\frac{5}{7}} \left( \frac{1-q}{1+q}\right)\, V_3+\mathcal{O}(V^3_3)\\
H_{44}\equiv& -\frac{4}{9}\sqrt{\frac{10}{7}} \left(1-\frac{3q}{(1+q)^2}\right)V^2_4+\mathcal{O}(V^4_4)
\end{split}\ee
Combining Eqs.~\eqref{eq:SPA} and~\eqref{eq:Hlm} with Eq.~\eqref{eq:predictedwf}, we get effective equations for the amplitude of higher harmonics in the detector frame as 
\be\begin{split}
\left|\frac{h^\mathrm{det}_{33}(3f)}{h^\mathrm{det}_{22}(2f)}\right| \simeq& \frac{3\sqrt{3}}{4\sqrt{2}}\, \left[\frac{1-q}{1+q}\right]\, (2\pi M_\mathrm{tot} f)^{1/3}\, \sin \iota \\
\left|\frac{h^\mathrm{det}_{44}(4f)}{h^\mathrm{det}_{22}(2f)}\right| \simeq& \frac{2\sqrt{2}}{3}\, \left[1-\frac{3 q}{(1+q)^2}\right](2\pi M_\mathrm{tot} f)^{2/3}\, \sin^2 \iota.
\end{split}
\label{eq:AmpFormulae}\ee

We show the performance of these equations in Fig.~\ref{fig:AmpModel} for a particular set of binary parameters and $\iota=\pi/3$.
We see that, although these amplitude formulae are constructed from low-order PN expressions, they are a fairly good approximation to the amplitude even in the quasi-linear regime (a similar behavior was also noted by Ref.~\cite{Boh20_HM_PN} while comparing PN expressions to time-domain numerical relativity waveforms). It is clear that maximizing over the amplitude of the HM waveform leads to maximization over inclination and masses, and the frequency dependence is independent of these waveforms.

In our case, each template in a particular bank has the same amplitude profile.
Following Eq.~\eqref{eq:AmpFormulae}, one could use $A^\mathrm{bank}_{33} (f) \propto  f^{1/3} A^\mathrm{bank}_{22}\left(2f/3\right)$ and $A^\mathrm{bank}_{44} (f) \propto f^{2/3} A^\mathrm{bank}_{22}\left(2f/4\right)$. 
We however use the following instead, as we find that this slightly improves the model accuracy in the non-linear regime:
\be
A^\mathrm{bank}_{33}(f) = \langle \overline{A_{33}}\rangle_\mathrm{bank} (f) \ ;\ A^\mathrm{bank}_{44}(f) = \langle \overline{A_{44}}\rangle_\mathrm{bank} (f)
\label{eq:HM_amplitudes}\ee
where the averaging is done over the normalized waveform samples in the bank, i.e., each sample has $\langle 
\overline{A_{ii}}| \overline{A_{ii}}\rangle=1$. After averaging, we normalize the bank amplitudes such that $\langle A^\mathrm{bank}_{\ell m}|A^\mathrm{bank}_{\ell m}\rangle=1$. It is worth mentioning that for modeling HM, we choose different frequency cutoffs than the ones used for 22: $f_\mathrm{min}(\ell \ell) = \frac{\ell}{2} f_\mathrm{min} (22)$, and a similar relation for $f_\mathrm{max}(\ell \ell)$.
%

\subsection{Model for phases}
\label{sec:HM_phases}

In our template banks, we model $h_{\ell m}$ as $A_{\ell m} e^{i\Psi_{\ell m}}$, where $A_{\ell m}$ is real. 
Let us start by writing the phase of the waveform in Eq.~\eqref{eq:SPA}:
\be
\arg h_{\ell m}(f) = m\Psi_\mathrm{SPA} \left(\bigg[ \frac{2\pi M f}{k}\bigg]^{1/3}\right) +\frac{\pi}{4} +\arg H_{\ell m} (f)
\ee

Ignoring the constant offset for the moment, one can see that a good approximation in the early inspiral regime is $\Psi_{\ell m}(f) \thickapprox \frac{m}{2} \Psi_{22} \big(\frac{2f}{m}\big)$. 
 However, it is important to mention that $H_{\ell m}$ can have imaginary terms in the higher order PN expansion which make $\arg H_{\ell m}(f)\neq 0$ (see Appendix~E of \cite{Gar20}). Furthermore, as we go into the non-linear regime, numerical relativity (NR) simulations show that
there is an additional offset (compared to PN) to the above approximation (see Fig.~5 of \cite{Gar20}). We therefore attempt to model all these frequency dependent offsets by calculating
\be
\Delta \Psi_{\ell m}(f) = \Psi_{\ell m}(f)-\frac{m}{2} \Psi_{22}\left(\frac{2f}{m}\right)
\label{eq:HM_phase_residuals}
\ee
directly from \texttt{IMRPhenomXHM} waveforms for each of the parameter samples used to create the template banks.   In Fig.~\ref{fig:PhaseModel} we show an example of a high mass-ratio case.
In the plot, we set $\Delta \Psi$ to be 0 at 100 Hz to make it easier to visually compare the frequency-dependent offset. We see that in some cases, the deviation can go up to a few radians and thus the above approximation is not sufficient for modeling $\Psi_{\ell m}$.

To remedy this, we try to model $\Delta \Psi_{\ell m}$ from \texttt{IMRPhenomXHM} as a function of intrinsic parameters. First, we decompose 
\be
\Delta \Psi_{33}(f)  = \langle\Delta \Psi_{33}\rangle_\mathrm{subbank} (f) + \sum_{i=0}^{2} c^{(33)}_i \Psi^\mathrm{SVD}_i(f) 
\label{eq:HM_SVD}
\ee
where $\langle \Delta \Psi \rangle(f)$ denotes the average over physical waveforms in the particular sub-bank. Next, we predict the coefficient 
\be
c^{(33)}_i = \mathrm{RF}_i (c_0^{(22)}, c_1^{(22)})
\label{eq:HM_calpha}
\ee
by training a random forest regressor (RF). Note that this approach is similar to the one used earlier in Section~\ref{sec:22_phases} to model the phases of the 22 waveforms. The RF takes the coefficients of a 22 template as input and outputs the first three $c_i^{(33)}$ coefficients (we checked empirically that including more than three coefficients had negligible effect on the results). Our final model for the phases becomes
\be\begin{split}
\Psi^\mathrm{model}_{33}(f) =  \frac{3}{2}& \Psi_{22}\left(\frac{2f}{3}\right) + \langle\Delta \Psi_{33}\rangle_\mathrm{subbank} (f)\\
 +& \sum_{i=0}^{2} \mathrm{RF}_i (c_0^{(22)}, c_1^{(22)}) \Psi^\mathrm{SVD}_i(f) 
\label{eq:33_phase_model}\end{split}\ee
We show the results from our model in the dashed line in Fig.~\ref{fig:PhaseModel}, where we indeed see an improvement compared to the naive approximation (see also Fig.~\ref{fig:PhaseModel_HM_apx}).
In the future, if needed, the accuracy of our phase model can be further improved by measuring $c^{(33)}_0$ instead of predicting it with RF. This would entail an additional dimension corresponding to the 33 phase, on top of the two existing dimensions corresponding to the 22 phase in our template banks. We leave further discussion of this point to Appendix~\ref{apx:HM_phase_improve}.
Finally, it is worth mentioning that we discussed the formalism for the 33 case in this sub-section, but an analogous method also applies for the 44 case.

\subsection{Model for amplitude ratios}
\label{sec:amp_ratios}

Now that we have modeled the phase profiles $\Psi^\mathrm{model} (f)$ and the normalized amplitude profiles $A^\mathrm{bank} (f)$ of the 22, 33 and 44 modes, the last ingredient in our template bank is the prior on the expected SNR in the HM relative to 22. To model this prior, we store in our subbanks a set of samples corresponding to the ratios:
\be
R_{\ell m} \equiv \frac{\langle h^\mathrm{XHM}_{\ell m}(f)|h^\mathrm{XHM}_{\ell m}(f)\rangle^{1/2}}{\langle h^\mathrm{XHM}_{22}(f)|h^\mathrm{XHM}_{22}(f)\rangle^{1/2}}
\label{eq:R33}\ee
which are calculated at the binary inclination $\iota=\pi/2$.
For generating the $R_{\ell m}$ samples, we use binary mass and spin parameters corresponding to the subbank from the astrophysical prior distribution given in section~\ref{sec:wf_samples}. Currently, we assume each template in a given subbank has the same set of mode SNR-ratio samples. This is sub-optimal if there is a significant variation in $R_{\ell\ell}$ values within the subbank parameter range. A better approach here could be to use template-dependent $R_{\ell\ell}$ samples generated using smoothing algorithms like kernel density estimation or normalizing flows, we leave exploration of this to an upcoming work. 

The model of the full waveforms from our template bank is therefore given by
\be\begin{split}
h(f)=&\, A^\mathrm{bank}_{22}(f)\, e^{i[\Psi^\mathrm{model}_{22}(f)+2\phi_0]}\\
&+R_{33} \sin \iota\, A^\mathrm{bank}_{33}(f)\, e^{i[\Psi^\mathrm{model}_{33}(f)+3\phi_0]} \\
&+R_{44} \sin^2 \iota\, A^\mathrm{bank}_{44}(f)\, e^{i[\Psi^\mathrm{model}_{44}(f)+4\phi_0]}  
\end{split}
\ee
where we have introduced the $\sin \iota$ terms following from Eq.~\eqref{eq:predictedwf} (note the $h^\mathrm{physical}$ waveforms we used in Eq.~\eqref{eq:R33} were simulated for the case of $\iota=\pi/2$ and $\phi_0=0$). An overall amplitude and the phase $\phi_0$ will be maximized over when we perform matched-filtering with the data (for a detailed derivation see Appendix A.2 of \cite{Wad23_Pipeline}).

\section{Overview of matched filtering with the banks}
\label{sec:matched_filtering}

In this section, we give a brief overview of the matched-filtering procedure with our banks and encourage the reader to refer to our companion paper (Ref.~\cite{Wad23_Pipeline}) for details. As shown in Fig.~\ref{fig:Triggering_modes}, we matched-filter each of the harmonics separately with the data and collect the complex SNR timeseries:
\be \begin{split}
\rho_{\ell\ell} (t) &= \langle \h_{\ell\ell}(f)\ |\ d(f)\ e^{i2\pi f t}\rangle
\label{eq:SNR_timeseries}
\end{split}\ee
where $\mathbb{h}_{\ell\ell} (f)\equiv A^\mathrm{bank}_{\ell\ell}(f) e^{i\Psi^\mathrm{model}_{\ell \ell} (f)}$ is the normalized template waveform and $d(f)$ is the data in the Fourier domain. 
Note that if we filter the modes with Gaussian noise, one would find some triggers with $|\rho_{33}/\rho_{22}|$ and $|\rho_{44}/\rho_{22}|$ values much larger than one (as $|\rho_{22}|^2, |\rho_{33}|^2$, and $|\rho_{44}|^2$ separately follow $\chi^2$ distributions). However, for astrophysical signals, the values of $|\rho_{33}/\rho_{22}|$ and $|\rho_{44}/\rho_{22}|$ are bounded (e.g., using the physical waveforms associated with the bank BBH-4, we find $|\rho_{33}/\rho_{22}|<0.78$ and $|\rho_{44}/\rho_{22}|<0.34$). In this section, our goal is to combine the different $\rho_{\ell\ell}$ in an optimal way which exploits this difference in their behavior under the signal and noise hypotheses. We use the Neyman-Pearson lemma \cite{neymanpearson}, which states that the optimal detection statistic is the ratio of evidence under the signal and the noise hypothesis \cite{Ols22_ias_o3a}: 
\be\begin{split}
&\frac{P(d|\mathcal{S})}{P(d|\mathcal{N})} = \frac{\int {\rm d}\Pi(\varepsilon_i,\varepsilon_e) P(\varepsilon_i,\varepsilon_e) \exp \left[ - \langle d-h | d-h\rangle/2 \right]}{\exp \left[-\langle d | d \rangle/2\right]}\\
& \ = \int {\rm d}\Pi(\varepsilon_i,\varepsilon_e) P(\varepsilon_i,\varepsilon_e) \exp \left[ \mathrm{Re}(\langle d | h \rangle) - \frac{1}{2}\langle h | h \rangle \right]
\label{eq:CoherentScore}
\end{split}\ee

where $d$ is the data and $h$ is the model waveform as a function of the intrinsic and extrinsic parameters $(\varepsilon_i,\varepsilon_e)$. $P$ and  ${\rm d}\Pi$ denotes our prior density and phase space differential respectively.
Note that Eq.~\eqref{eq:CoherentScore} assumes that the noise is Gaussian but we relax that assumption later in Ref.~\cite{Wad23_Pipeline}.
In the matched-filtering step of the pipeline, we maximize over the intrinsic parameters instead of marginalizing them. The inner products in the above equations can be written for an individual detector as
\be \begin{split}
\langle d | h \rangle \propto  e^{2i\phi_0} \rho_{22} + e^{3i\phi_0} R_{33} \sin \iota\, \rho_{33} + e^{4i\phi_0} R_{44} \sin^2 \iota\, \rho_{44}
\end{split} \ee
and 
\be
\langle h (f) | h (f) \rangle = \sum_{\ell, \ell'} \langle
  h_{\ell \ell}(f) | h_{\ell'\ell'}(f) \rangle \propto \sum_{\ell, \ell'}  e^{i\phi_0 (\ell'-\ell)}\,  \mathbb{C}_{\ell,\ell'}
\ee
where the covariance matrix between the modes is given by
\be \begin{split}
\mathbb{C}  =& \begin{bmatrix}
1 & e^{i\phi_0} R_{33} \langle\h_{22}|\h_{33}\rangle & e^{2i\phi_0} R_{44} \langle\h_{22}|\h_{44}\rangle\\
- & R_{33}^2 \sin^2 \iota & e^{i\phi_0} R_{33}R_{44} \langle\h_{33}|\h_{44}\rangle\\
- & - & R_{44}^2 \sin^4 \iota
\end{bmatrix}
\end{split}\ee
(the lower-triangular elements are not shown but can be obtained by Hermitian conjugation). An approximate version of the marginalized detection statistic therefore becomes
\be\begin{split}
\exp\left(\frac{\rho^2_\mathrm{marg}}{2}\right) \simeq & \int {\rm d}\Pi(\iota, \phi_0, R_{33}, R_{44})\\
&\ \ \times \exp \left[ \mathrm{Re}(\langle d | h \rangle) - \frac{1}{2}\langle h | h \rangle \right]
\label{eq:CoherentScore2}
\end{split}\ee
One could also maximize over the parameters instead of marginalizing in case one wants a less optimal but less computationally-expensive detection statistic. To further speed-up the calculation, one could use the Gram-Schmidt orthogonalization method discussed in the next section (which makes the $\langle h_m | h_{m'} \rangle$ matrix diagonal). For marginalizing over $R_{33}$ and $R_{44}$, we use the list of samples simulated in Sec.~\ref{sec:amp_ratios}.
We show the performance of this detection statistic in Ref.~\cite{Wad23_Pipeline} and leave further discussion of the matched-filtering and trigger-ranking process of our pipeline to that paper.
%


\section{Calculating effectualness of banks}
\label{sec:Effectualness}

In this section, we calculate the effectualness of our HM banks.
We replace the data in Eq.~\eqref{eq:SNR_timeseries} with normalized test waveforms $h^\mathrm{XHM}$ from \texttt{IMRPhenomXHM} which include all the currently available modes: (2,2), (3,3), (4,4), (2,1), (3,2). Then, using our templates, we obtain the complex match timeseries for each mode in our bank: $\rho_{22} (t), \rho_{33} (t), \rho_{44} (t)$. For choosing the inclinations of the test waveforms, we use the probability distribution from Eq.~\eqref{eq:p_iota}, which is derived from the brightness of the 22 waveform.
Typically the 22 mode is louder than HM, so we choose the inclination distribution based on it. We expect minor changes to our results if the inclination distribution is based on the brightness of the full waveform including HM. The masses and spins corresponding to the test waveforms are sampled from the astrophysical prior distribution given in section~\ref{sec:wf_samples}.

Note that $\mathbb{h}_{22}$, $\mathbb{h}_{33}$, $\mathbb{h}_{44}$ in Eq.~\eqref{eq:SNR_timeseries} are nearly (but not exactly) orthogonal to each other. The off-diagonal terms of their covariance matrix $C_{\ell \ell'} \equiv \langle \mathbb{h}_{\ell\ell}\, (f)|\mathbb{h}_{\ell'\ell'}\, (f)\rangle$ are typically $\lesssim 0.1$, but at times can be larger (e.g., in the case of short duration waveforms for particular cases of PSD profiles).
To calculate the match, one could use the inner product $(\sum_{\ell\ell'}\rho^*_{\ell\ell} [C^{-1}]_{\ell \ell'} \rho_{\ell'\ell'})^{1/2}$.

\begin{figure}
\centering
\includegraphics[width=0.48\textwidth,keepaspectratio=true]{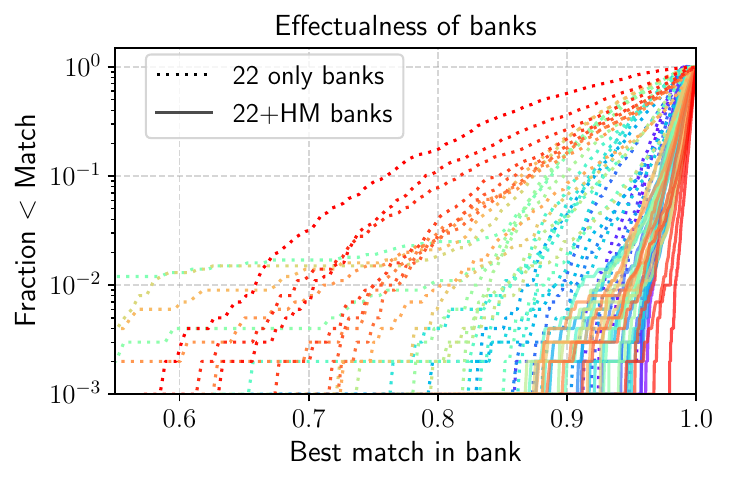}
\includegraphics[width=0.48\textwidth,keepaspectratio=true]{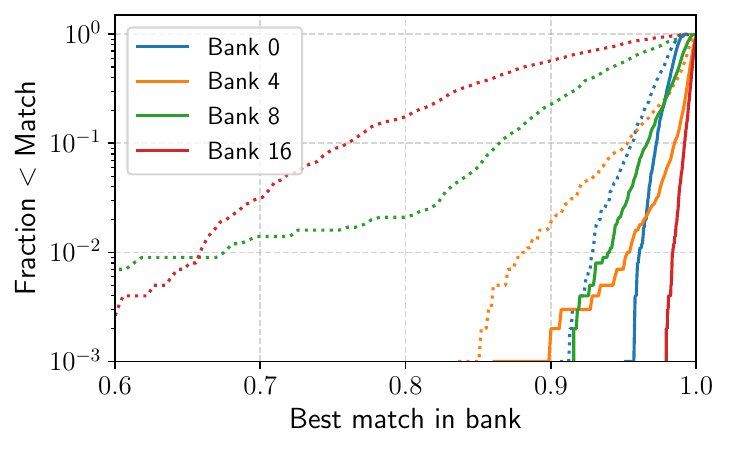}
\caption{
Effectualness of our template banks tested on random waveforms drawn from \texttt{IMRPhenomXHM} including the modes (2,2), (3,3), (4,4), (2,1), (3,2), (4,3). The templates are drawn from mass and spin priors given in Eq.~\eqref{eq:prior}. The $y$-axis shows the fraction of the test samples with match less than the value on the $x$-axis (also referred to as the template bank fitting factor (FF)).
\textbf{Top:} Results for all banks. The color varies in ascending order with the banks (sub-banks of BBH-0 are in blue, while sub-banks of BBH-16 are in red). \textbf{Bottom:} We show the same effectualness again, but for only a few banks in increasing order to emphasize that the improvement upon including HM is more significant for high-mass banks.}
\label{fig:Effectualness}
\end{figure}

To make the calculation easier, we orthogonalize the harmonics using the Gram-Schmidt method: we keep 22 fixed and orthogonalize 33 with respect to 22, and then orthogonalize 44 with respect to 22 and the orthogonalized 33.
We then normalize the harmonics such that  $ \langle \mathbb{h}^\perp_{\ell\ell}\, (f)|\mathbb{h}^\perp_{\ell'\ell'}\, (f)\rangle = \delta^K_{\ell \ell'}$, where $\delta^K$ is the Kronecker delta function. Therefore, we get
 \be
\mathrm{match}^2= \max_{t}\Big(|\rho_{22}(t)|^2+|\rho_{33}(t)|^2+|\rho_{44}(t)|^2\Big)
\label{eq:Match}
\ee
where $\rho_{\ell\ell}(t) = \langle h^\mathrm{XHM}|\h^\perp_{\ell\ell}\, e^{i2\pi f t} \rangle/\sqrt{\langle h^\mathrm{XHM}|h^\mathrm{XHM}\rangle}$. We show the results in Fig.~\ref{fig:Effectualness}. Note that the matches of different modes are combined at the same time (we do not allow an unphysical time shift between the different modes).
Overall, we obtain matches $\gtrsim 0.95$ for more than 99\% of the test waveforms. In the future, if needed, one could further improve upon the effectualness values for some of the high mass banks with HM by having more than one 33 template corresponding to individual 22 templates, as discussed in Appendix~\ref{apx:HM_phase_improve}. This will increase the number of templates but, as we see from Table~\ref{tab:Ntemplates}, the number of templates in the high-mass region is already very low and therefore the overall computational cost would only increase marginally. 

For calculating the effectualness, in principle, we would have to compute the inner product of the test waveform against each waveform in the bank to find the best match. To save computational effort, we follow a method similar to Ref.~\citetalias{ias_template_bank_PSD_roulet2019} where we use the advantage offered by our SVD-based method for modeling the phases.
For each test waveform, we first find the closest grid point in the $c^{(22)}_0, c^{(22)}_1$ space by unwrapping the phase of the 22 component of the waveform and then taking the dot product with the SVD phase basis waveforms from Eq.~\eqref{eq:phases_22}. We then construct the 22, 33, 44 modes corresponding to that grid point using Eqs.~\eqref{eq:HM_SVD} and \eqref{eq:HM_calpha}, and then calculate the match. In principle, one could take the inner product of the test waveform with all the templates in the bank, which would find a maximum match greater than or equal to our method's result, and hence our method is conservative.

\section{Discussion}
\label{sec:discussion}

%

\subsection{Application of our method to stochastic placement template banks }
\label{sec:stochastic}
The (2,2) templates in our HM banks were placed using the geometric placement algorithm. Note however that our method of triggering the data separately with different harmonics
is agnostic to the (2,2) template placement method. Our only requirement is that every (2,2) template in the bank is accompanied by normalized (3,3) and (4,4) templates.

Most of the other (2,2) template banks used in the literature use the stochastic placement algorithm.
In this case, waveforms are drawn randomly from the desired parameter space, and 
the bank gradually converges to the prescribed effectualness. Over-coverage is limited
by only accepting newly drawn waveforms that differ sufficiently from the ones already in the bank (rejecting those that are too similar to at least one existing template).

One of the advantages of geometric placement algorithms
is that the template density over the parameter space is optimal and no region is over-covered. One of the drawbacks of the geometric placement method we used is that the templates based on coefficients from SVD cannot be interpreted directly in terms of intuitive physical parameters such as masses and spins. Although the geometric placement model is optimal in the low chirp mass region, for the high-mass region the number of templates is small (see Table~\ref{tab:Ntemplates}) and stochastic placement could also be convenient. Since the next stages of the IAS search pipeline (triggering with data, coincidence, and candidate ranking) are currently written for the case of geometric placement, we used that over the full mass regime in our template banks.

We now outline a procedure for
creating a stochastic placement template bank that includes HM. Similar to our method in Section~\ref{sec:amp_ratios}, a template would correspond to a set of normalized waveforms for each mode $\h_{22}, \h_{33}, \h_{44}$ and a list of amplitude ratio samples $(R^{(i)}_{33}, R^{(i)}_{44})$. To generate the bank, draw a set of intrinsic parameters. Simulate physical waveforms for the three modes $h_{22}, h_{33}, h_{44}$ using the parameters $\phi_\mathrm{0}=0$ and $\iota=\pi/2$.
For each generated waveform, calculate the $R_{33}, R_{44}$ ratios given in Eq.~\eqref{eq:R33} and then normalize the templates.

If the match of the simulated $\h_{22}$ with the 22 waveforms already present in the bank is worse than a particular threshold, add the set of waveforms to the bank and assign $(R_{33}$, $R_{44})$ to the new template.  If the 22 match is above the threshold, discard the waveforms but assign ($R_{33}$, $R_{44}$) to the pre-existing template which gave the best 22 match\footnote{It is worth mentioning that, in case one wants to improve the bank effectualness even further, one could store multiple $\h_{33}, \h_{44}$ associated to each $\h_{22}$ by checking if the mismatch between the simulated $\h_{33}, \h_{44}$ and the $\h_{33}, \h_{44}$ already present in the bank is larger than a particular threshold.}. The next step of matched-filtering the data with individual modes and combining the SNR timeseries remains the same as in our search (see Ref.~\cite{Wad23_Pipeline}).


\section{Conclusions and future prospects}
\label{sec:conclusions}

We have presented a new efficient method for creating template banks that include higher-order modes (HM) such as 33, 44 in addition to the dominant 22 mode for aligned-spin waveforms. We do not combine the different harmonics
at the template level, which requires a new template for each sampled value of inclination and initial orbital phase. Instead, we store the 22, 33, 44 harmonics separately and only combine their matched-filter timeseries.
We model the amplitudes and phases separately for the three modes using post-Newtonian formulae and machine learning methods. We show that our method is effectual by testing it with waveforms from \texttt{IMRPhenomXHM}.

In principle, this mode-by-mode filtering method can also be used to reduce the computational burden of other searches 
aiming 
to include additional degrees of freedom in the template banks. The 
key is being
able to decompose waveforms made with complex physical effects into a dominant nearly-circular 22 mode and a few sister waveforms (which include the additional degrees of freedom, similar to the 33, 44 waveforms in our case). In a recent paper \cite{McI23}, the authors used a similar approach of decomposing precessing waveforms into different modes and matched-filtering them separately with the data. 
Including the effect of eccentricity on GW waveforms
is another physically interesting template extension that adds
extra degrees of freedom to the banks, and we plan to explore this case in a future study.


In our companion papers, we discuss our detection statistic \cite{Wad23_Pipeline}, and report on the results of our HM search of the O3 LIGO--Virgo data with our new template bank \cite{Wad23_HM_Events}.

The template banks described in this article are publicly available alongside the \texttt{IAS-HM} pipeline at \url{https://github.com/JayWadekar/gwIAS-HM}. The notebooks: \texttt{1.Template\_banks.ipynb} and \texttt{2.Astrophysical\_prior.ipynb} can be used to make our template banks from scratch, and also include code snippets to reproduce the plots in this paper.

\acknowledgments

We thank Horng Sheng Chia, Sophia Yi, Hang Yu, Tom Edwards, Ajith Parameswaran, Mukesh Singh, Liang Dai, and Aaron Zimmerman for helpful discussions. 
DW gratefully acknowledges support from the Friends of the Institute for Advanced Study Membership and the Keck Foundation. 
TV acknowledges support from NSF grants 2012086 and 2309360, the Alfred P. Sloan Foundation through grant number FG-2023-20470, the BSF through award number 2022136 and the Hellman Family Faculty Fellowship. BZ is supported by the Israel Science Foundation, NSF-BSF and by a research grant from the Willner Family Leadership Institute for the Weizmann Institute of Science. MZ is supported by NSF 2209991 and NSF-BSF 2207583. This research was also supported in part by the National Science Foundation under Grant No. NSF PHY-1748958. We also thank KITP and ICTS-TIFR for their hospitality during the completion of a part of this work. 

This research has made use of data, software and/or web tools obtained from the Gravitational Wave Open Science Center (\url{https://www.gw-openscience.org/}), a service of LIGO Laboratory, the LIGO Scientific Collaboration and the Virgo Collaboration. LIGO Laboratory and Advanced LIGO are funded by the United States National Science Foundation (NSF) as well as the Science and Technology Facilities Council (STFC) of the United Kingdom, the Max-Planck-Society (MPS), and the State of Niedersachsen/Germany for support of the construction of Advanced LIGO and construction and operation of the GEO600 detector. Additional support for Advanced LIGO was provided by the Australian Research Council. Virgo is funded, through the European Gravitational Observatory (EGO), by the French Centre National de Recherche Scientifique (CNRS), the Italian Istituto Nazionale di Fisica Nucleare (INFN) and the Dutch Nikhef, with contributions by institutions from Belgium, Germany, Greece, Hungary, Ireland, Japan, Monaco, Poland, Portugal, Spain.

\appendix

\begin{figure}
\centering
\includegraphics[width=0.48\textwidth,keepaspectratio=true]{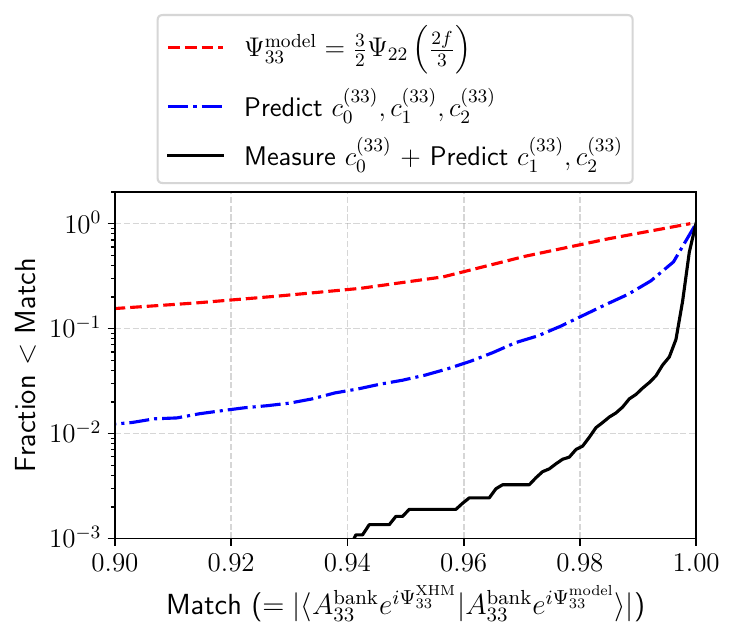}
\includegraphics[width=0.48\textwidth,keepaspectratio=true]{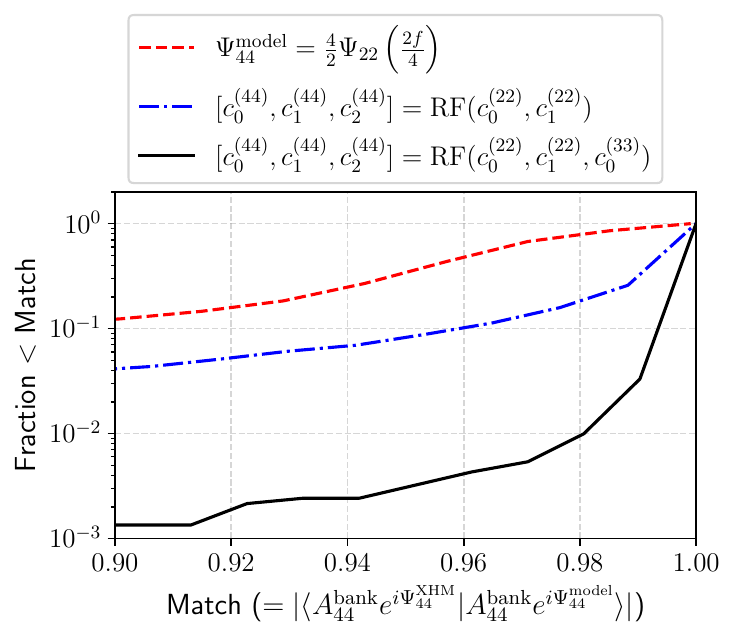}
\caption{Similar to Fig.~\ref{fig:RF_calpha}, but showing the accuracy of different models of the 33 (44) phases instead of the 22 phases in the top (bottom) panel. In red, we show the naive model derived from early-inspiral PN expansion. For the blue and black curves, we use the model in Eq.~\eqref{eq:33_model_apx} but predict the SVD coefficients using a random forest regressor with different sets of inputs. In blue, we show the predictions only using the 22 coefficients as input (which is the default model used for the effectualness results shown in the the main text). To further improve the results in the future, one could measure $c_0^{(33)}$ by having a few 33 templates in the bank corresponding to a single 22 template, and this would give an improved phase model shown in black.}
\label{fig:PhaseModel_HM_apx}
\end{figure}

\section{Polarizations of HM waveforms}
\label{apx:Polarizations}

For a binary merger which does not exhibit precession, the waveform can be expressed in the frequency domain, using the stationary-phase approximation (SPA) as \cite{Mil21}:
\be\begin{split}
h_+(f) &= \frac{D_0}{D_L} \sum_{\ell \geq 2}\sum_{m=0}^{\ell} A_{\ell m}^+(\iota) e^{im\phi_0} h_{\ell m}(f)\\
h_\times(f) &= \frac{D_0}{D_L} \sum_{\ell \geq 2}\sum_{m=0}^{\ell} A_{\ell m}^\times(\iota)\, i e^{im\phi_0} h_{\ell m}(f)
\label{eq:h_plus_cross}
\end{split}\ee
where the $A_{\ell m}$ factors are functions of only the inclination angle and are given by \cite{Mil21}
\be\begin{split}
A^+_{22}&=\frac{1+\cos^2\iota}{2};\ \ A^\times_{22}=\cos \iota\\
(A_{33}^+&,A_{33}^\times) = 2\sin \iota\, (A_{22}^+,A_{22}^\times)\\
(A_{44}^+&,A_{44}^\times) = 2\sin^2 \iota\,  (A_{22}^+,A_{22}^\times)\\
A^+_{21}&=\sin \iota;\ \ A^\times_{21}=\sin \iota \cos \iota\\
A^+_{32}&=1-2\cos^2 \iota;\ \ A^\times_{32}=\frac{1}{2} (\cos \iota - 3 \cos^3 \iota)\\
\label{eq:Aij}
\end{split}\ee

Note that for the special case of $\ell=m$ waveforms, $h_\times (f)/h_+(f) = 2i\cos\iota/(1+\cos^2\iota)$. However this ratio has a different dependence on $\iota$ for the 21 and 32 cases. Writing an analogue of Eq.~\eqref{eq:predictedwf}, but also including 21 and 32, we get
\be \begin{split}\label{eq:predictedwf2}
  h_\mathrm{det}(f) =\, &\frac{D_0}{D_L} \left[ F_{+} \frac{1 + \cos^2\iota}{2} \, + \, i\cos \iota\, F_{\times}\right]
   \bigg[ e^{2 i \phi_0} \,  h_{22}(f)\\ 
   &+\, 2 \sin \iota\, e^{3 i \phi_0}\,  h_{33}(f)+\, 2\sin^2 \iota\, e^{4 i \phi_0}\, h_{44}(f)\bigg]\\
   +& \frac{D_0}{D_L}[ F_{+} \, + \, i\cos \iota\, F_{\times}] [\sin \iota\, e^{i \phi_0} h_{21}(f)]\\
   +&  \frac{D_0}{D_L}\bigg[ F_{+} (1-2\cos^2 \iota) \, + \, i \frac{F_{\times}}{2}(\cos \iota - 3 \cos^3 \iota)\bigg]\\
   & \times \bigg[e^{2i \phi_0} h_{32}(f) \bigg]
\end{split}\ee

Unlike Eq.~\eqref{eq:predictedwf}, now the detector responses cannot be factored out as an overall normalization term. One would therefore need to simultaneously maximize/marginalize over detector responses, $\phi_0$ and $\iota$ to get an optimal SNR statistic. Note that previous studies including $\ell \neq m$ modes have introduced techniques to maximize over detector responses \cite{Har18, Cha22}. We leave including 21 and 32 modes in our analysis method to future work and only consider the 22, 33, 44 modes in this study.

\section{Improving effectualness of our HM banks}
\label{apx:HM_phase_improve}

In Eq.~\eqref{eq:33_phase_model} of Section~\ref{sec:HM_phases}, we had discussed our model for the HM phases as
\be
\Psi^\mathrm{model}_{\ell\ell}(f)  = \frac{\ell}{2}\Psi_{22}\left(\frac{2f}{\ell}\right) + \langle\Delta \Psi_{\ell\ell}\rangle_\mathrm{bank} (f) + \sum_{i=1}^{3} c^{(\ell\ell)}_i \Psi^\mathrm{SVD}_i(f) 
\label{eq:33_model_apx}\ee
where the coefficients were predicted using a random forest regressor (RF) 
\be
[c^{(\ell\ell)}_0, c^{(\ell\ell)}_1, c^{(\ell\ell)}_2] = \mathrm{RF} (c_0^{(22)}, c_1^{(22)})
\ee

We show the accuracy of this approximation in the blue dot-dashed line in Fig.~\ref{fig:PhaseModel_HM_apx}, where we show the match given by $|\langle A^\mathrm{bank}_{\ell\ell}(f) e^{i\Psi_{\ell\ell}(f)}| A^\mathrm{bank}_{\ell\ell}(f) e^{i\Psi^\mathrm{model}_{\ell\ell}(f)} \rangle|$ of the phases from our model against the phases obtained from the simulated IMRPhenomXHM waveforms.
In this Appendix, we discuss a possible way of improving effectualness results from the high-mass banks in Fig.~\ref{fig:Effectualness}. Instead of predicting $c^{(33)}_0$, one would need to measure it directly from the data by introducing multiple 33 templates corresponding to a grid in the $c^{(33)}_0$ values (in this case, corresponding to a single 22 template, we would have a few 33 templates in the bank). The next coefficients can be predicted using RF in a way similar to earlier as
\be
[c^{(33)}_1, c^{(33)}_2] = \mathrm{RF} (c_0^{(22)}, c_1^{(22)}, c^{(33)}_0)
\ee
Note that we would not however need any additional templates corresponding to the 44 mode, because the SVD coefficients for 44 can be obtained using
\be
[c^{(44)}_0, c^{(44)}_1, c^{(44)}_2] = \mathrm{RF} (c_0^{(22)}, c_1^{(22)}, c^{(33)}_0)
\ee
We show the improvement from using the previous two equations in the black line in Fig.~\ref{fig:PhaseModel_HM_apx}.

\bibliographystyle{apsrev4-1}
\bibliography{HM}

\begin{thebibliography}{56}%
\makeatletter
\providecommand \@ifxundefined [1]{%
 \@ifx{#1\undefined}
}%
\providecommand \@ifnum [1]{%
 \ifnum #1\expandafter \@firstoftwo
 \else \expandafter \@secondoftwo
 \fi
}%
\providecommand \@ifx [1]{%
 \ifx #1\expandafter \@firstoftwo
 \else \expandafter \@secondoftwo
 \fi
}%
\providecommand \natexlab [1]{#1}%
\providecommand \enquote  [1]{``#1''}%
\providecommand \bibnamefont  [1]{#1}%
\providecommand \bibfnamefont [1]{#1}%
\providecommand \citenamefont [1]{#1}%
\providecommand \href@noop [0]{\@secondoftwo}%
\providecommand \href [0]{\begingroup \@sanitize@url \@href}%
\providecommand \@href[1]{\@@startlink{#1}\@@href}%
\providecommand \@@href[1]{\endgroup#1\@@endlink}%
\providecommand \@sanitize@url [0]{\catcode `\\12\catcode `\$12\catcode
  `\&12\catcode `\#12\catcode `\^12\catcode `\_12\catcode `\%12\relax}%
\providecommand \@@startlink[1]{}%
\providecommand \@@endlink[0]{}%
\providecommand \url  [0]{\begingroup\@sanitize@url \@url }%
\providecommand \@url [1]{\endgroup\@href {#1}{\urlprefix }}%
\providecommand \urlprefix  [0]{URL }%
\providecommand \Eprint [0]{\href }%
\providecommand \doibase [0]{http://dx.doi.org/}%
\providecommand \selectlanguage [0]{\@gobble}%
\providecommand \bibinfo  [0]{\@secondoftwo}%
\providecommand \bibfield  [0]{\@secondoftwo}%
\providecommand \translation [1]{[#1]}%
\providecommand \BibitemOpen [0]{}%
\providecommand \bibitemStop [0]{}%
\providecommand \bibitemNoStop [0]{.\EOS\space}%
\providecommand \EOS [0]{\spacefactor3000\relax}%
\providecommand \BibitemShut  [1]{\csname bibitem#1\endcsname}%
\let\auto@bib@innerbib\@empty
\bibitem [{\citenamefont {{Wadekar}}\ \emph {et~al.}(2023)\citenamefont
  {{Wadekar}}, \citenamefont {{Roulet}}, \citenamefont {{Venumadhav}},
  \citenamefont {{Mehta}}, \citenamefont {{Zackay}}, \citenamefont {{Mushkin}},
  \citenamefont {{Olsen}},\ and\ \citenamefont
  {{Zaldarriaga}}}]{Wad23_HM_Events}%
  \BibitemOpen
  \bibfield  {author} {\bibinfo {author} {\bibfnamefont {D.}~\bibnamefont
  {{Wadekar}}}, \bibinfo {author} {\bibfnamefont {J.}~\bibnamefont {{Roulet}}},
  \bibinfo {author} {\bibfnamefont {T.}~\bibnamefont {{Venumadhav}}}, \bibinfo
  {author} {\bibfnamefont {A.~K.}\ \bibnamefont {{Mehta}}}, \bibinfo {author}
  {\bibfnamefont {B.}~\bibnamefont {{Zackay}}}, \bibinfo {author}
  {\bibfnamefont {J.}~\bibnamefont {{Mushkin}}}, \bibinfo {author}
  {\bibfnamefont {S.}~\bibnamefont {{Olsen}}}, \ and\ \bibinfo {author}
  {\bibfnamefont {M.}~\bibnamefont {{Zaldarriaga}}},\ }\href@noop {} {\bibfield
   {journal} {\bibinfo  {journal} {arXiv e-prints}\ ,\ \bibinfo {eid}
  {arXiv:2312.06631}} (\bibinfo {year} {2023})},\ \Eprint
  {http://arxiv.org/abs/2312.06631} {arXiv:2312.06631 [gr-qc]} \BibitemShut
  {NoStop}%
\bibitem [{\citenamefont {{Wadekar}}\ \emph {et~al.}(2024)\citenamefont
  {{Wadekar}}, \citenamefont {{Venumadhav}}, \citenamefont {{Roulet}},
  \citenamefont {{Mehta}}, \citenamefont {{Zackay}}, \citenamefont
  {{Mushkin}},\ and\ \citenamefont {{Zaldarriaga}}}]{Wad23_Pipeline}%
  \BibitemOpen
  \bibfield  {author} {\bibinfo {author} {\bibfnamefont {D.}~\bibnamefont
  {{Wadekar}}}, \bibinfo {author} {\bibfnamefont {T.}~\bibnamefont
  {{Venumadhav}}}, \bibinfo {author} {\bibfnamefont {J.}~\bibnamefont
  {{Roulet}}}, \bibinfo {author} {\bibfnamefont {A.~K.}\ \bibnamefont
  {{Mehta}}}, \bibinfo {author} {\bibfnamefont {B.}~\bibnamefont {{Zackay}}},
  \bibinfo {author} {\bibfnamefont {J.}~\bibnamefont {{Mushkin}}}, \ and\
  \bibinfo {author} {\bibfnamefont {M.}~\bibnamefont {{Zaldarriaga}}},\ }\href
  {\doibase 10.1103/PhysRevD.110.044063} {\bibfield  {journal} {\bibinfo
  {journal} {\prd}\ }\textbf {\bibinfo {volume} {110}},\ \bibinfo {eid}
  {044063} (\bibinfo {year} {2024})},\ \Eprint
  {http://arxiv.org/abs/2405.17400} {arXiv:2405.17400 [gr-qc]} \BibitemShut
  {NoStop}%
\bibitem [{\citenamefont {{Chandra}}\ \emph {et~al.}(2022)\citenamefont
  {{Chandra}}, \citenamefont {{Calder{\'o}n Bustillo}}, \citenamefont {{Pai}},\
  and\ \citenamefont {{Harry}}}]{Cha22}%
  \BibitemOpen
  \bibfield  {author} {\bibinfo {author} {\bibfnamefont {K.}~\bibnamefont
  {{Chandra}}}, \bibinfo {author} {\bibfnamefont {J.}~\bibnamefont
  {{Calder{\'o}n Bustillo}}}, \bibinfo {author} {\bibfnamefont
  {A.}~\bibnamefont {{Pai}}}, \ and\ \bibinfo {author} {\bibfnamefont
  {I.}~\bibnamefont {{Harry}}},\ }\href@noop {} {\bibfield  {journal} {\bibinfo
   {journal} {arXiv e-prints}\ ,\ \bibinfo {eid} {arXiv:2207.01654}} (\bibinfo
  {year} {2022})},\ \Eprint {http://arxiv.org/abs/2207.01654} {arXiv:2207.01654
  [gr-qc]} \BibitemShut {NoStop}%
\bibitem [{\citenamefont {{Schmidt}}\ \emph {et~al.}(2023)\citenamefont
  {{Schmidt}}, \citenamefont {{Gadre}},\ and\ \citenamefont
  {{Caudill}}}]{Sch23_NF_TemplateBank}%
  \BibitemOpen
  \bibfield  {author} {\bibinfo {author} {\bibfnamefont {S.}~\bibnamefont
  {{Schmidt}}}, \bibinfo {author} {\bibfnamefont {B.}~\bibnamefont {{Gadre}}},
  \ and\ \bibinfo {author} {\bibfnamefont {S.}~\bibnamefont {{Caudill}}},\
  }\href {\doibase 10.48550/arXiv.2302.00436} {\bibfield  {journal} {\bibinfo
  {journal} {arXiv e-prints}\ ,\ \bibinfo {eid} {arXiv:2302.00436}} (\bibinfo
  {year} {2023})},\ \Eprint {http://arxiv.org/abs/2302.00436} {arXiv:2302.00436
  [gr-qc]} \BibitemShut {NoStop}%
\bibitem [{\citenamefont {Harry}\ \emph {et~al.}(2018)\citenamefont {Harry},
  \citenamefont {Bustillo},\ and\ \citenamefont {Nitz}}]{Har18}%
  \BibitemOpen
  \bibfield  {author} {\bibinfo {author} {\bibfnamefont {I.}~\bibnamefont
  {Harry}}, \bibinfo {author} {\bibfnamefont {J.~C.}\ \bibnamefont {Bustillo}},
  \ and\ \bibinfo {author} {\bibfnamefont {A.}~\bibnamefont {Nitz}},\ }\href
  {\doibase 10.1103/PhysRevD.97.023004} {\bibfield  {journal} {\bibinfo
  {journal} {Phys. Rev. D}\ }\textbf {\bibinfo {volume} {97}},\ \bibinfo
  {pages} {023004} (\bibinfo {year} {2018})}\BibitemShut {NoStop}%
\bibitem [{\citenamefont {Abbott}\ \emph {et~al.}(2016)\citenamefont {Abbott}
  \emph {et~al.}}]{O1catalog_LVC2016}%
  \BibitemOpen
  \bibfield  {author} {\bibinfo {author} {\bibfnamefont {B.~P.}\ \bibnamefont
  {Abbott}} \emph {et~al.} (\bibinfo {collaboration} {LIGO Scientific
  Collaboration and Virgo Collaboration}),\ }\href {\doibase
  10.1103/PhysRevX.6.041015} {\bibfield  {journal} {\bibinfo  {journal} {Phys.
  Rev. X}\ }\textbf {\bibinfo {volume} {6}},\ \bibinfo {pages} {041015}
  (\bibinfo {year} {2016})}\BibitemShut {NoStop}%
\bibitem [{\citenamefont {Abbott}\ \emph {et~al.}(2019)\citenamefont {Abbott}
  \emph {et~al.}}]{gwtc1_o2catalog_LVC2018}%
  \BibitemOpen
  \bibfield  {author} {\bibinfo {author} {\bibfnamefont {B.~P.}\ \bibnamefont
  {Abbott}} \emph {et~al.} (\bibinfo {collaboration} {LIGO Scientific,
  Virgo}),\ }\href {\doibase 10.1103/PhysRevX.9.031040} {\bibfield  {journal}
  {\bibinfo  {journal} {Phys. Rev. X}\ }\textbf {\bibinfo {volume} {9}},\
  \bibinfo {pages} {031040} (\bibinfo {year} {2019})},\ \Eprint
  {http://arxiv.org/abs/1811.12907} {arXiv:1811.12907 [astro-ph.HE]}
  \BibitemShut {NoStop}%
\bibitem [{\citenamefont {{Abbott}}\ \emph {et~al.}(2021)\citenamefont
  {{Abbott}} \emph {et~al.}}]{lvc_o3a_gwtc2_catalog_2021}%
  \BibitemOpen
  \bibfield  {author} {\bibinfo {author} {\bibfnamefont {R.}~\bibnamefont
  {{Abbott}}} \emph {et~al.},\ }\href {\doibase 10.1103/PhysRevX.11.021053}
  {\bibfield  {journal} {\bibinfo  {journal} {Physical Review X}\ }\textbf
  {\bibinfo {volume} {11}},\ \bibinfo {eid} {021053} (\bibinfo {year}
  {2021})},\ \Eprint {http://arxiv.org/abs/2010.14527} {arXiv:2010.14527
  [gr-qc]} \BibitemShut {NoStop}%
\bibitem [{\citenamefont {Sachdev}\ \emph {et~al.}(2019)\citenamefont
  {Sachdev}, \citenamefont {Caudill} \emph {et~al.}}]{gstlal}%
  \BibitemOpen
  \bibfield  {author} {\bibinfo {author} {\bibfnamefont {S.}~\bibnamefont
  {Sachdev}}, \bibinfo {author} {\bibfnamefont {S.}~\bibnamefont {Caudill}},
  \emph {et~al.},\ }\href {http://arxiv.org/abs/1901.08580v1} {\  (\bibinfo
  {year} {2019})},\ \Eprint {http://arxiv.org/abs/1901.08580v1}
  {arXiv:1901.08580v1 [gr-qc]} \BibitemShut {NoStop}%
\bibitem [{\citenamefont {{Usman}}\ \emph {et~al.}(2016)\citenamefont
  {{Usman}}, \citenamefont {{Nitz}}, \citenamefont {{Harry}} \emph
  {et~al.}}]{PYCBCPipeline}%
  \BibitemOpen
  \bibfield  {author} {\bibinfo {author} {\bibfnamefont {S.~A.}\ \bibnamefont
  {{Usman}}}, \bibinfo {author} {\bibfnamefont {A.~H.}\ \bibnamefont {{Nitz}}},
  \bibinfo {author} {\bibfnamefont {I.~W.}\ \bibnamefont {{Harry}}},  \emph
  {et~al.},\ }\href {\doibase 10.1088/0264-9381/33/21/215004} {\bibfield
  {journal} {\bibinfo  {journal} {Classical and Quantum Gravity}\ }\textbf
  {\bibinfo {volume} {33}},\ \bibinfo {eid} {215004} (\bibinfo {year}
  {2016})},\ \Eprint {http://arxiv.org/abs/1508.02357} {arXiv:1508.02357
  [gr-qc]} \BibitemShut {NoStop}%
\bibitem [{\citenamefont {Andres}\ \emph {et~al.}(2022)\citenamefont {Andres},
  \citenamefont {Assiduo}, \citenamefont {Aubin}, \citenamefont {Chierici},
  \citenamefont {Estevez}, \citenamefont {Faedi}, \citenamefont {Guidi},
  \citenamefont {Juste}, \citenamefont {Marion}, \citenamefont {Mours},
  \citenamefont {Nitoglia},\ and\ \citenamefont
  {Sordini}}]{mbta_o3a_pastro_andres2022}%
  \BibitemOpen
  \bibfield  {author} {\bibinfo {author} {\bibfnamefont {N.}~\bibnamefont
  {Andres}}, \bibinfo {author} {\bibfnamefont {M.}~\bibnamefont {Assiduo}},
  \bibinfo {author} {\bibfnamefont {F.}~\bibnamefont {Aubin}}, \bibinfo
  {author} {\bibfnamefont {R.}~\bibnamefont {Chierici}}, \bibinfo {author}
  {\bibfnamefont {D.}~\bibnamefont {Estevez}}, \bibinfo {author} {\bibfnamefont
  {F.}~\bibnamefont {Faedi}}, \bibinfo {author} {\bibfnamefont {G.~M.}\
  \bibnamefont {Guidi}}, \bibinfo {author} {\bibfnamefont {V.}~\bibnamefont
  {Juste}}, \bibinfo {author} {\bibfnamefont {F.}~\bibnamefont {Marion}},
  \bibinfo {author} {\bibfnamefont {B.}~\bibnamefont {Mours}}, \bibinfo
  {author} {\bibfnamefont {E.}~\bibnamefont {Nitoglia}}, \ and\ \bibinfo
  {author} {\bibfnamefont {V.}~\bibnamefont {Sordini}},\ }\href {\doibase
  10.1088/1361-6382/ac482a} {\bibfield  {journal} {\bibinfo  {journal}
  {Classical and Quantum Gravity}\ }\textbf {\bibinfo {volume} {39}},\ \bibinfo
  {pages} {055002} (\bibinfo {year} {2022})}\BibitemShut {NoStop}%
\bibitem [{\citenamefont {{The LIGO Scientific Collaboration}}\ \emph
  {et~al.}(2021)\citenamefont {{The LIGO Scientific Collaboration}},
  \citenamefont {{the Virgo Collaboration}} \emph
  {et~al.}}]{lvc_o3a_deep_gwtc2_1_update_2021}%
  \BibitemOpen
  \bibfield  {author} {\bibinfo {author} {\bibnamefont {{The LIGO Scientific
  Collaboration}}}, \bibinfo {author} {\bibnamefont {{the Virgo
  Collaboration}}},  \emph {et~al.},\ }\href
  {https://ui.adsabs.harvard.edu/abs/2021arXiv210801045T} {\bibfield  {journal}
  {\bibinfo  {journal} {arXiv e-prints}\ ,\ \bibinfo {eid} {arXiv:2108.01045}}
  (\bibinfo {year} {2021})},\ \Eprint {http://arxiv.org/abs/2108.01045}
  {arXiv:2108.01045 [gr-qc]} \BibitemShut {NoStop}%
\bibitem [{\citenamefont {Abbott}\ \emph {et~al.}(2021)\citenamefont {Abbott}
  \emph {et~al.}}]{lvc_gwtc3_o3_ab_catalog_2021}%
  \BibitemOpen
  \bibfield  {author} {\bibinfo {author} {\bibfnamefont {R.}~\bibnamefont
  {Abbott}} \emph {et~al.} (\bibinfo {collaboration} {LIGO Scientific, VIRGO,
  KAGRA}),\ }\href@noop {} {\  (\bibinfo {year} {2021})},\ \Eprint
  {http://arxiv.org/abs/2111.03606} {arXiv:2111.03606 [gr-qc]} \BibitemShut
  {NoStop}%
\bibitem [{\citenamefont {{The LIGO Scientific Collaboration}}\ \emph
  {et~al.}(2024)\citenamefont {{The LIGO Scientific Collaboration}},
  \citenamefont {{the Virgo Collaboration}},\ and\ \citenamefont {{the KAGRA
  Collaboration}}}]{GW230529}%
  \BibitemOpen
  \bibfield  {author} {\bibinfo {author} {\bibnamefont {{The LIGO Scientific
  Collaboration}}}, \bibinfo {author} {\bibnamefont {{the Virgo
  Collaboration}}}, \ and\ \bibinfo {author} {\bibnamefont {{the KAGRA
  Collaboration}}},\ }\href {\doibase 10.48550/arXiv.2404.04248} {\bibfield
  {journal} {\bibinfo  {journal} {arXiv e-prints}\ ,\ \bibinfo {eid}
  {arXiv:2404.04248}} (\bibinfo {year} {2024})},\ \Eprint
  {http://arxiv.org/abs/2404.04248} {arXiv:2404.04248 [astro-ph.HE]}
  \BibitemShut {NoStop}%
\bibitem [{\citenamefont {{Venumadhav}}\ \emph {et~al.}(2020)\citenamefont
  {{Venumadhav}}, \citenamefont {{Zackay}}, \citenamefont {{Roulet}},
  \citenamefont {{Dai}},\ and\ \citenamefont
  {{Zaldarriaga}}}]{ias_o2_pipeline_new_events_prd2020}%
  \BibitemOpen
  \bibfield  {author} {\bibinfo {author} {\bibfnamefont {T.}~\bibnamefont
  {{Venumadhav}}}, \bibinfo {author} {\bibfnamefont {B.}~\bibnamefont
  {{Zackay}}}, \bibinfo {author} {\bibfnamefont {J.}~\bibnamefont {{Roulet}}},
  \bibinfo {author} {\bibfnamefont {L.}~\bibnamefont {{Dai}}}, \ and\ \bibinfo
  {author} {\bibfnamefont {M.}~\bibnamefont {{Zaldarriaga}}},\ }\href {\doibase
  10.1103/PhysRevD.101.083030} {\bibfield  {journal} {\bibinfo  {journal}
  {\prd}\ }\textbf {\bibinfo {volume} {101}},\ \bibinfo {eid} {083030}
  (\bibinfo {year} {2020})},\ \Eprint {http://arxiv.org/abs/1904.07214}
  {arXiv:1904.07214 [astro-ph.HE]} \BibitemShut {NoStop}%
\bibitem [{\citenamefont {Olsen}\ \emph {et~al.}(2022)\citenamefont {Olsen},
  \citenamefont {Venumadhav}, \citenamefont {Mushkin}, \citenamefont {Roulet},
  \citenamefont {Zackay},\ and\ \citenamefont {Zaldarriaga}}]{Ols22_ias_o3a}%
  \BibitemOpen
  \bibfield  {author} {\bibinfo {author} {\bibfnamefont {S.}~\bibnamefont
  {Olsen}}, \bibinfo {author} {\bibfnamefont {T.}~\bibnamefont {Venumadhav}},
  \bibinfo {author} {\bibfnamefont {J.}~\bibnamefont {Mushkin}}, \bibinfo
  {author} {\bibfnamefont {J.}~\bibnamefont {Roulet}}, \bibinfo {author}
  {\bibfnamefont {B.}~\bibnamefont {Zackay}}, \ and\ \bibinfo {author}
  {\bibfnamefont {M.}~\bibnamefont {Zaldarriaga}},\ }\href {\doibase
  10.1103/PhysRevD.106.043009} {\bibfield  {journal} {\bibinfo  {journal}
  {Phys. Rev. D}\ }\textbf {\bibinfo {volume} {106}},\ \bibinfo {pages}
  {043009} (\bibinfo {year} {2022})}\BibitemShut {NoStop}%
\bibitem [{\citenamefont {{Mehta}}\ \emph {et~al.}(2023)\citenamefont
  {{Mehta}}, \citenamefont {{Olsen}}, \citenamefont {{Wadekar}}, \citenamefont
  {{Roulet}}, \citenamefont {{Venumadhav}}, \citenamefont {{Mushkin}},
  \citenamefont {{Zackay}},\ and\ \citenamefont
  {{Zaldarriaga}}}]{Meh23_ias_o3b}%
  \BibitemOpen
  \bibfield  {author} {\bibinfo {author} {\bibfnamefont {A.~K.}\ \bibnamefont
  {{Mehta}}}, \bibinfo {author} {\bibfnamefont {S.}~\bibnamefont {{Olsen}}},
  \bibinfo {author} {\bibfnamefont {D.}~\bibnamefont {{Wadekar}}}, \bibinfo
  {author} {\bibfnamefont {J.}~\bibnamefont {{Roulet}}}, \bibinfo {author}
  {\bibfnamefont {T.}~\bibnamefont {{Venumadhav}}}, \bibinfo {author}
  {\bibfnamefont {J.}~\bibnamefont {{Mushkin}}}, \bibinfo {author}
  {\bibfnamefont {B.}~\bibnamefont {{Zackay}}}, \ and\ \bibinfo {author}
  {\bibfnamefont {M.}~\bibnamefont {{Zaldarriaga}}},\ }\href {\doibase
  10.48550/arXiv.2311.06061} {\bibfield  {journal} {\bibinfo  {journal} {arXiv
  e-prints}\ ,\ \bibinfo {eid} {arXiv:2311.06061}} (\bibinfo {year} {2023})},\
  \Eprint {http://arxiv.org/abs/2311.06061} {arXiv:2311.06061 [gr-qc]}
  \BibitemShut {NoStop}%
\bibitem [{\citenamefont {{Chia}}\ \emph {et~al.}(2023)\citenamefont {{Chia}},
  \citenamefont {{Edwards}}, \citenamefont {{Wadekar}}, \citenamefont
  {{Zimmerman}}, \citenamefont {{Olsen}}, \citenamefont {{Roulet}},
  \citenamefont {{Venumadhav}}, \citenamefont {{Zackay}},\ and\ \citenamefont
  {{Zaldarriaga}}}]{Chi23}%
  \BibitemOpen
  \bibfield  {author} {\bibinfo {author} {\bibfnamefont {H.~S.}\ \bibnamefont
  {{Chia}}}, \bibinfo {author} {\bibfnamefont {T.~D.~P.}\ \bibnamefont
  {{Edwards}}}, \bibinfo {author} {\bibfnamefont {D.}~\bibnamefont
  {{Wadekar}}}, \bibinfo {author} {\bibfnamefont {A.}~\bibnamefont
  {{Zimmerman}}}, \bibinfo {author} {\bibfnamefont {S.}~\bibnamefont
  {{Olsen}}}, \bibinfo {author} {\bibfnamefont {J.}~\bibnamefont {{Roulet}}},
  \bibinfo {author} {\bibfnamefont {T.}~\bibnamefont {{Venumadhav}}}, \bibinfo
  {author} {\bibfnamefont {B.}~\bibnamefont {{Zackay}}}, \ and\ \bibinfo
  {author} {\bibfnamefont {M.}~\bibnamefont {{Zaldarriaga}}},\ }\href {\doibase
  10.48550/arXiv.2306.00050} {\bibfield  {journal} {\bibinfo  {journal} {arXiv
  e-prints}\ ,\ \bibinfo {eid} {arXiv:2306.00050}} (\bibinfo {year} {2023})},\
  \Eprint {http://arxiv.org/abs/2306.00050} {arXiv:2306.00050 [gr-qc]}
  \BibitemShut {NoStop}%
\bibitem [{\citenamefont {Nitz}\ \emph {et~al.}(2020)\citenamefont {Nitz},
  \citenamefont {Dent}, \citenamefont {Davies}, \citenamefont {Kumar},
  \citenamefont {Capano}, \citenamefont {Harry}, \citenamefont {Mozzon},
  \citenamefont {Nuttall}, \citenamefont {Lundgren},\ and\ \citenamefont
  {T\'apai}}]{NitzCatalog_2-OGC_o2_2020}%
  \BibitemOpen
  \bibfield  {author} {\bibinfo {author} {\bibfnamefont {A.~H.}\ \bibnamefont
  {Nitz}}, \bibinfo {author} {\bibfnamefont {T.}~\bibnamefont {Dent}}, \bibinfo
  {author} {\bibfnamefont {G.~S.}\ \bibnamefont {Davies}}, \bibinfo {author}
  {\bibfnamefont {S.}~\bibnamefont {Kumar}}, \bibinfo {author} {\bibfnamefont
  {C.~D.}\ \bibnamefont {Capano}}, \bibinfo {author} {\bibfnamefont
  {I.}~\bibnamefont {Harry}}, \bibinfo {author} {\bibfnamefont
  {S.}~\bibnamefont {Mozzon}}, \bibinfo {author} {\bibfnamefont
  {L.}~\bibnamefont {Nuttall}}, \bibinfo {author} {\bibfnamefont
  {A.}~\bibnamefont {Lundgren}}, \ and\ \bibinfo {author} {\bibfnamefont
  {M.}~\bibnamefont {T\'apai}},\ }\href {\doibase 10.3847/1538-4357/ab733f}
  {\bibfield  {journal} {\bibinfo  {journal} {Astrophys. J.}\ }\textbf
  {\bibinfo {volume} {891}},\ \bibinfo {pages} {123} (\bibinfo {year}
  {2020})},\ \Eprint {http://arxiv.org/abs/1910.05331} {arXiv:1910.05331
  [astro-ph.HE]} \BibitemShut {NoStop}%
\bibitem [{\citenamefont {{Nitz}}\ \emph {et~al.}(2021)\citenamefont {{Nitz}},
  \citenamefont {{Capano}}, \citenamefont {{Kumar}}, \citenamefont {{Wang}},
  \citenamefont {{Kastha}}, \citenamefont {{Sch{\"a}fer}}, \citenamefont
  {{Dhurkunde}},\ and\ \citenamefont {{Cabero}}}]{nitz_o3a_3ogc_catalog_2021}%
  \BibitemOpen
  \bibfield  {author} {\bibinfo {author} {\bibfnamefont {A.~H.}\ \bibnamefont
  {{Nitz}}}, \bibinfo {author} {\bibfnamefont {C.~D.}\ \bibnamefont
  {{Capano}}}, \bibinfo {author} {\bibfnamefont {S.}~\bibnamefont {{Kumar}}},
  \bibinfo {author} {\bibfnamefont {Y.-F.}\ \bibnamefont {{Wang}}}, \bibinfo
  {author} {\bibfnamefont {S.}~\bibnamefont {{Kastha}}}, \bibinfo {author}
  {\bibfnamefont {M.}~\bibnamefont {{Sch{\"a}fer}}}, \bibinfo {author}
  {\bibfnamefont {R.}~\bibnamefont {{Dhurkunde}}}, \ and\ \bibinfo {author}
  {\bibfnamefont {M.}~\bibnamefont {{Cabero}}},\ }\href
  {https://ui.adsabs.harvard.edu/abs/2021arXiv210509151N} {\bibfield  {journal}
  {\bibinfo  {journal} {arXiv e-prints}\ ,\ \bibinfo {eid} {arXiv:2105.09151}}
  (\bibinfo {year} {2021})},\ \Eprint {http://arxiv.org/abs/2105.09151}
  {arXiv:2105.09151 [astro-ph.HE]} \BibitemShut {NoStop}%
\bibitem [{\citenamefont {Nitz}\ \emph {et~al.}(2021)\citenamefont {Nitz},
  \citenamefont {Kumar}, \citenamefont {Wang}, \citenamefont {Kastha},
  \citenamefont {Wu}, \citenamefont {Schäfer}, \citenamefont {Dhurkunde},\
  and\ \citenamefont {Capano}}]{nitz_4ogc_o3_ab_catalog_2021}%
  \BibitemOpen
  \bibfield  {author} {\bibinfo {author} {\bibfnamefont {A.~H.}\ \bibnamefont
  {Nitz}}, \bibinfo {author} {\bibfnamefont {S.}~\bibnamefont {Kumar}},
  \bibinfo {author} {\bibfnamefont {Y.-F.}\ \bibnamefont {Wang}}, \bibinfo
  {author} {\bibfnamefont {S.}~\bibnamefont {Kastha}}, \bibinfo {author}
  {\bibfnamefont {S.}~\bibnamefont {Wu}}, \bibinfo {author} {\bibfnamefont
  {M.}~\bibnamefont {Schäfer}}, \bibinfo {author} {\bibfnamefont
  {R.}~\bibnamefont {Dhurkunde}}, \ and\ \bibinfo {author} {\bibfnamefont
  {C.~D.}\ \bibnamefont {Capano}},\ }\href@noop {} {\enquote {\bibinfo {title}
  {4-{OGC}: Catalog of gravitational waves from compact-binary mergers},}\ }
  (\bibinfo {year} {2021}),\ \Eprint {http://arxiv.org/abs/2112.06878}
  {arXiv:2112.06878 [astro-ph.HE]} \BibitemShut {NoStop}%
\bibitem [{\citenamefont {Thorne}(1980)}]{Tho80}%
  \BibitemOpen
  \bibfield  {author} {\bibinfo {author} {\bibfnamefont {K.~S.}\ \bibnamefont
  {Thorne}},\ }\href {\doibase 10.1103/RevModPhys.52.299} {\bibfield  {journal}
  {\bibinfo  {journal} {Rev. Mod. Phys.}\ }\textbf {\bibinfo {volume} {52}},\
  \bibinfo {pages} {299} (\bibinfo {year} {1980})}\BibitemShut {NoStop}%
\bibitem [{\citenamefont {{Varma}}\ \emph {et~al.}(2014)\citenamefont
  {{Varma}}, \citenamefont {{Ajith}}, \citenamefont {{Husa}}, \citenamefont
  {{Bustillo}}, \citenamefont {{Hannam}},\ and\ \citenamefont
  {{P{\"u}rrer}}}]{Var14}%
  \BibitemOpen
  \bibfield  {author} {\bibinfo {author} {\bibfnamefont {V.}~\bibnamefont
  {{Varma}}}, \bibinfo {author} {\bibfnamefont {P.}~\bibnamefont {{Ajith}}},
  \bibinfo {author} {\bibfnamefont {S.}~\bibnamefont {{Husa}}}, \bibinfo
  {author} {\bibfnamefont {J.~C.}\ \bibnamefont {{Bustillo}}}, \bibinfo
  {author} {\bibfnamefont {M.}~\bibnamefont {{Hannam}}}, \ and\ \bibinfo
  {author} {\bibfnamefont {M.}~\bibnamefont {{P{\"u}rrer}}},\ }\href {\doibase
  10.1103/PhysRevD.90.124004} {\bibfield  {journal} {\bibinfo  {journal}
  {\prd}\ }\textbf {\bibinfo {volume} {90}},\ \bibinfo {eid} {124004} (\bibinfo
  {year} {2014})},\ \Eprint {http://arxiv.org/abs/1409.2349} {arXiv:1409.2349
  [gr-qc]} \BibitemShut {NoStop}%
\bibitem [{\citenamefont {{Sakon}}\ \emph {et~al.}(2022)\citenamefont
  {{Sakon}}, \citenamefont {{Tsukada}} \emph
  {et~al.}}]{Sak23_TemplateBank_GSTLAL}%
  \BibitemOpen
  \bibfield  {author} {\bibinfo {author} {\bibfnamefont {S.}~\bibnamefont
  {{Sakon}}}, \bibinfo {author} {\bibfnamefont {L.}~\bibnamefont {{Tsukada}}},
  \emph {et~al.},\ }\href {\doibase 10.48550/arXiv.2211.16674} {\bibfield
  {journal} {\bibinfo  {journal} {arXiv e-prints}\ ,\ \bibinfo {eid}
  {arXiv:2211.16674}} (\bibinfo {year} {2022})},\ \Eprint
  {http://arxiv.org/abs/2211.16674} {arXiv:2211.16674 [gr-qc]} \BibitemShut
  {NoStop}%
\bibitem [{\citenamefont {Pekowsky}\ \emph {et~al.}(2013)\citenamefont
  {Pekowsky}, \citenamefont {Healy}, \citenamefont {Shoemaker},\ and\
  \citenamefont {Laguna}}]{HMeffect_ParameterSpaceDependency_PekowskyPRD2013}%
  \BibitemOpen
  \bibfield  {author} {\bibinfo {author} {\bibfnamefont {L.}~\bibnamefont
  {Pekowsky}}, \bibinfo {author} {\bibfnamefont {J.}~\bibnamefont {Healy}},
  \bibinfo {author} {\bibfnamefont {D.}~\bibnamefont {Shoemaker}}, \ and\
  \bibinfo {author} {\bibfnamefont {P.}~\bibnamefont {Laguna}},\ }\href
  {\doibase 10.1103/physrevd.87.084008} {\bibfield  {journal} {\bibinfo
  {journal} {Physical Review D}\ }\textbf {\bibinfo {volume} {87}} (\bibinfo
  {year} {2013}),\ 10.1103/physrevd.87.084008}\BibitemShut {NoStop}%
\bibitem [{\citenamefont {Healy}\ \emph {et~al.}(2013)\citenamefont {Healy},
  \citenamefont {Laguna}, \citenamefont {Pekowsky},\ and\ \citenamefont
  {Shoemaker}}]{HMeffect_RelativeModeSignificance_HealyPRD2013}%
  \BibitemOpen
  \bibfield  {author} {\bibinfo {author} {\bibfnamefont {J.}~\bibnamefont
  {Healy}}, \bibinfo {author} {\bibfnamefont {P.}~\bibnamefont {Laguna}},
  \bibinfo {author} {\bibfnamefont {L.}~\bibnamefont {Pekowsky}}, \ and\
  \bibinfo {author} {\bibfnamefont {D.}~\bibnamefont {Shoemaker}},\ }\href
  {\doibase 10.1103/physrevd.88.024034} {\bibfield  {journal} {\bibinfo
  {journal} {Physical Review D}\ }\textbf {\bibinfo {volume} {88}} (\bibinfo
  {year} {2013}),\ 10.1103/physrevd.88.024034}\BibitemShut {NoStop}%
\bibitem [{\citenamefont {Capano}\ \emph {et~al.}(2014)\citenamefont {Capano},
  \citenamefont {Pan},\ and\ \citenamefont {Buonanno}}]{Cap14}%
  \BibitemOpen
  \bibfield  {author} {\bibinfo {author} {\bibfnamefont {C.}~\bibnamefont
  {Capano}}, \bibinfo {author} {\bibfnamefont {Y.}~\bibnamefont {Pan}}, \ and\
  \bibinfo {author} {\bibfnamefont {A.}~\bibnamefont {Buonanno}},\ }\href
  {\doibase 10.1103/physrevd.89.102003} {\bibfield  {journal} {\bibinfo
  {journal} {Physical Review D}\ }\textbf {\bibinfo {volume} {89}} (\bibinfo
  {year} {2014}),\ 10.1103/physrevd.89.102003}\BibitemShut {NoStop}%
\bibitem [{\citenamefont {Bustillo}\ \emph {et~al.}(2016)\citenamefont
  {Bustillo}, \citenamefont {Husa}, \citenamefont {Sintes},\ and\ \citenamefont
  {Pürrer}}]{HMeffect_AlignedSearchImpactCalderonBustilloPRD2016}%
  \BibitemOpen
  \bibfield  {author} {\bibinfo {author} {\bibfnamefont {J.~C.}\ \bibnamefont
  {Bustillo}}, \bibinfo {author} {\bibfnamefont {S.}~\bibnamefont {Husa}},
  \bibinfo {author} {\bibfnamefont {A.~M.}\ \bibnamefont {Sintes}}, \ and\
  \bibinfo {author} {\bibfnamefont {M.}~\bibnamefont {Pürrer}},\ }\href
  {\doibase 10.1103/physrevd.93.084019} {\bibfield  {journal} {\bibinfo
  {journal} {Physical Review D}\ }\textbf {\bibinfo {volume} {93}} (\bibinfo
  {year} {2016}),\ 10.1103/physrevd.93.084019}\BibitemShut {NoStop}%
\bibitem [{\citenamefont {Bustillo}\ \emph {et~al.}(2017)\citenamefont
  {Bustillo}, \citenamefont {Laguna},\ and\ \citenamefont
  {Shoemaker}}]{HMandPrecessionEffect_HeavySearchImpact_CalderonBustilloPRD2017}%
  \BibitemOpen
  \bibfield  {author} {\bibinfo {author} {\bibfnamefont {J.~C.}\ \bibnamefont
  {Bustillo}}, \bibinfo {author} {\bibfnamefont {P.}~\bibnamefont {Laguna}}, \
  and\ \bibinfo {author} {\bibfnamefont {D.}~\bibnamefont {Shoemaker}},\ }\href
  {\doibase 10.1103/physrevd.95.104038} {\bibfield  {journal} {\bibinfo
  {journal} {Physical Review D}\ }\textbf {\bibinfo {volume} {95}} (\bibinfo
  {year} {2017}),\ 10.1103/physrevd.95.104038}\BibitemShut {NoStop}%
\bibitem [{\citenamefont {Bustillo}\ \emph {et~al.}(2018)\citenamefont
  {Bustillo}, \citenamefont {Salemi}, \citenamefont {Canton},\ and\
  \citenamefont {Jani}}]{HMeffect_IMBHsearchImpact_CalderonBustilloPRD2018}%
  \BibitemOpen
  \bibfield  {author} {\bibinfo {author} {\bibfnamefont {J.~C.}\ \bibnamefont
  {Bustillo}}, \bibinfo {author} {\bibfnamefont {F.}~\bibnamefont {Salemi}},
  \bibinfo {author} {\bibfnamefont {T.~D.}\ \bibnamefont {Canton}}, \ and\
  \bibinfo {author} {\bibfnamefont {K.~P.}\ \bibnamefont {Jani}},\ }\href
  {\doibase 10.1103/physrevd.97.024016} {\bibfield  {journal} {\bibinfo
  {journal} {Physical Review D}\ }\textbf {\bibinfo {volume} {97}} (\bibinfo
  {year} {2018}),\ 10.1103/physrevd.97.024016}\BibitemShut {NoStop}%
\bibitem [{\citenamefont {{Mills}}\ and\ \citenamefont
  {{Fairhurst}}(2021)}]{Mil21}%
  \BibitemOpen
  \bibfield  {author} {\bibinfo {author} {\bibfnamefont {C.}~\bibnamefont
  {{Mills}}}\ and\ \bibinfo {author} {\bibfnamefont {S.}~\bibnamefont
  {{Fairhurst}}},\ }\href {\doibase 10.1103/PhysRevD.103.024042} {\bibfield
  {journal} {\bibinfo  {journal} {\prd}\ }\textbf {\bibinfo {volume} {103}},\
  \bibinfo {eid} {024042} (\bibinfo {year} {2021})},\ \Eprint
  {http://arxiv.org/abs/2007.04313} {arXiv:2007.04313 [gr-qc]} \BibitemShut
  {NoStop}%
\bibitem [{\citenamefont {{Sharma}}\ \emph {et~al.}(2022)\citenamefont
  {{Sharma}}, \citenamefont {{Chandra}},\ and\ \citenamefont {{Pai}}}]{Sha22}%
  \BibitemOpen
  \bibfield  {author} {\bibinfo {author} {\bibfnamefont {K.}~\bibnamefont
  {{Sharma}}}, \bibinfo {author} {\bibfnamefont {K.}~\bibnamefont {{Chandra}}},
  \ and\ \bibinfo {author} {\bibfnamefont {A.}~\bibnamefont {{Pai}}},\
  }\href@noop {} {\bibfield  {journal} {\bibinfo  {journal} {arXiv e-prints}\
  ,\ \bibinfo {eid} {arXiv:2208.02545}} (\bibinfo {year} {2022})},\ \Eprint
  {http://arxiv.org/abs/2208.02545} {arXiv:2208.02545 [astro-ph.HE]}
  \BibitemShut {NoStop}%
\bibitem [{\citenamefont {{Zhang}}\ \emph {et~al.}(2023)\citenamefont
  {{Zhang}}, \citenamefont {{Dai}},\ and\ \citenamefont {{Liang}}}]{Zha23}%
  \BibitemOpen
  \bibfield  {author} {\bibinfo {author} {\bibfnamefont {C.}~\bibnamefont
  {{Zhang}}}, \bibinfo {author} {\bibfnamefont {N.}~\bibnamefont {{Dai}}}, \
  and\ \bibinfo {author} {\bibfnamefont {D.}~\bibnamefont {{Liang}}},\ }\href
  {\doibase 10.1103/PhysRevD.108.044076} {\bibfield  {journal} {\bibinfo
  {journal} {\prd}\ }\textbf {\bibinfo {volume} {108}},\ \bibinfo {eid}
  {044076} (\bibinfo {year} {2023})},\ \Eprint
  {http://arxiv.org/abs/2306.13871} {arXiv:2306.13871 [gr-qc]} \BibitemShut
  {NoStop}%
\bibitem [{\citenamefont {{Abbott}}\ \emph
  {et~al.}(2020{\natexlab{a}})\citenamefont {{Abbott}} \emph
  {et~al.}}]{GW190814}%
  \BibitemOpen
  \bibfield  {author} {\bibinfo {author} {\bibfnamefont {R.}~\bibnamefont
  {{Abbott}}} \emph {et~al.},\ }\href {\doibase 10.3847/2041-8213/ab960f}
  {\bibfield  {journal} {\bibinfo  {journal} {\apjl}\ }\textbf {\bibinfo
  {volume} {896}},\ \bibinfo {eid} {L44} (\bibinfo {year}
  {2020}{\natexlab{a}})},\ \Eprint {http://arxiv.org/abs/2006.12611}
  {arXiv:2006.12611 [astro-ph.HE]} \BibitemShut {NoStop}%
\bibitem [{\citenamefont {{Abbott}}\ \emph
  {et~al.}(2020{\natexlab{b}})\citenamefont {{Abbott}}, \citenamefont {{LIGO
  Scientific Collaboration}},\ and\ \citenamefont {{Virgo
  Collaboration}}}]{GW190412}%
  \BibitemOpen
  \bibfield  {author} {\bibinfo {author} {\bibfnamefont {R.}~\bibnamefont
  {{Abbott}}}, \bibinfo {author} {\bibnamefont {{LIGO Scientific
  Collaboration}}}, \ and\ \bibinfo {author} {\bibnamefont {{Virgo
  Collaboration}}},\ }\href {\doibase 10.1103/PhysRevD.102.043015} {\bibfield
  {journal} {\bibinfo  {journal} {\prd}\ }\textbf {\bibinfo {volume} {102}},\
  \bibinfo {eid} {043015} (\bibinfo {year} {2020}{\natexlab{b}})},\ \Eprint
  {http://arxiv.org/abs/2004.08342} {arXiv:2004.08342 [astro-ph.HE]}
  \BibitemShut {NoStop}%
\bibitem [{\citenamefont {{Sch{\"a}fer}}\ \emph {et~al.}(2022)\citenamefont
  {{Sch{\"a}fer}}, \citenamefont {{Zelenka}}, \citenamefont {{Nitz}},
  \citenamefont {{Wang}}, \citenamefont {{Wu}}, \citenamefont {{Guo}},
  \citenamefont {{Cao}}, \citenamefont {{Ren}}, \citenamefont {{Nousi}},
  \citenamefont {{Stergioulas}}, \citenamefont {{Iosif}}, \citenamefont
  {{Koloniari}}, \citenamefont {{Tefas}}, \citenamefont {{Passalis}},
  \citenamefont {{Salemi}}, \citenamefont {{Vedovato}}, \citenamefont
  {{Klimenko}}, \citenamefont {{Mishra}}, \citenamefont {{Br{\"u}gmann}},
  \citenamefont {{Cuoco}}, \citenamefont {{Huerta}}, \citenamefont
  {{Messenger}},\ and\ \citenamefont {{Ohme}}}]{Sch22}%
  \BibitemOpen
  \bibfield  {author} {\bibinfo {author} {\bibfnamefont {M.~B.}\ \bibnamefont
  {{Sch{\"a}fer}}}, \bibinfo {author} {\bibfnamefont {O.}~\bibnamefont
  {{Zelenka}}}, \bibinfo {author} {\bibfnamefont {A.~H.}\ \bibnamefont
  {{Nitz}}}, \bibinfo {author} {\bibfnamefont {H.}~\bibnamefont {{Wang}}},
  \bibinfo {author} {\bibfnamefont {S.}~\bibnamefont {{Wu}}}, \bibinfo {author}
  {\bibfnamefont {Z.-K.}\ \bibnamefont {{Guo}}}, \bibinfo {author}
  {\bibfnamefont {Z.}~\bibnamefont {{Cao}}}, \bibinfo {author} {\bibfnamefont
  {Z.}~\bibnamefont {{Ren}}}, \bibinfo {author} {\bibfnamefont
  {P.}~\bibnamefont {{Nousi}}}, \bibinfo {author} {\bibfnamefont
  {N.}~\bibnamefont {{Stergioulas}}}, \bibinfo {author} {\bibfnamefont
  {P.}~\bibnamefont {{Iosif}}}, \bibinfo {author} {\bibfnamefont {A.~E.}\
  \bibnamefont {{Koloniari}}}, \bibinfo {author} {\bibfnamefont
  {A.}~\bibnamefont {{Tefas}}}, \bibinfo {author} {\bibfnamefont
  {N.}~\bibnamefont {{Passalis}}}, \bibinfo {author} {\bibfnamefont
  {F.}~\bibnamefont {{Salemi}}}, \bibinfo {author} {\bibfnamefont
  {G.}~\bibnamefont {{Vedovato}}}, \bibinfo {author} {\bibfnamefont
  {S.}~\bibnamefont {{Klimenko}}}, \bibinfo {author} {\bibfnamefont
  {T.}~\bibnamefont {{Mishra}}}, \bibinfo {author} {\bibfnamefont
  {B.}~\bibnamefont {{Br{\"u}gmann}}}, \bibinfo {author} {\bibfnamefont
  {E.}~\bibnamefont {{Cuoco}}}, \bibinfo {author} {\bibfnamefont {E.~A.}\
  \bibnamefont {{Huerta}}}, \bibinfo {author} {\bibfnamefont {C.}~\bibnamefont
  {{Messenger}}}, \ and\ \bibinfo {author} {\bibfnamefont {F.}~\bibnamefont
  {{Ohme}}},\ }\href@noop {} {\bibfield  {journal} {\bibinfo  {journal} {arXiv
  e-prints}\ ,\ \bibinfo {eid} {arXiv:2209.11146}} (\bibinfo {year} {2022})},\
  \Eprint {http://arxiv.org/abs/2209.11146} {arXiv:2209.11146 [astro-ph.IM]}
  \BibitemShut {NoStop}%
\bibitem [{\citenamefont {{Tian}}\ \emph {et~al.}(2023)\citenamefont {{Tian}},
  \citenamefont {{Huerta}},\ and\ \citenamefont {{Zheng}}}]{Tin23}%
  \BibitemOpen
  \bibfield  {author} {\bibinfo {author} {\bibfnamefont {M.}~\bibnamefont
  {{Tian}}}, \bibinfo {author} {\bibfnamefont {E.~A.}\ \bibnamefont
  {{Huerta}}}, \ and\ \bibinfo {author} {\bibfnamefont {H.}~\bibnamefont
  {{Zheng}}},\ }\href {\doibase 10.48550/arXiv.2310.00052} {\bibfield
  {journal} {\bibinfo  {journal} {arXiv e-prints}\ ,\ \bibinfo {eid}
  {arXiv:2310.00052}} (\bibinfo {year} {2023})},\ \Eprint
  {http://arxiv.org/abs/2310.00052} {arXiv:2310.00052 [astro-ph.IM]}
  \BibitemShut {NoStop}%
\bibitem [{\citenamefont {{Garc{\'\i}a-Quir{\'o}s}}\ \emph
  {et~al.}(2020)\citenamefont {{Garc{\'\i}a-Quir{\'o}s}}, \citenamefont
  {{Colleoni}}, \citenamefont {{Husa}}, \citenamefont {{Estell{\'e}s}},
  \citenamefont {{Pratten}}, \citenamefont {{Ramos-Buades}}, \citenamefont
  {{Mateu-Lucena}},\ and\ \citenamefont {{Jaume}}}]{Gar20}%
  \BibitemOpen
  \bibfield  {author} {\bibinfo {author} {\bibfnamefont {C.}~\bibnamefont
  {{Garc{\'\i}a-Quir{\'o}s}}}, \bibinfo {author} {\bibfnamefont
  {M.}~\bibnamefont {{Colleoni}}}, \bibinfo {author} {\bibfnamefont
  {S.}~\bibnamefont {{Husa}}}, \bibinfo {author} {\bibfnamefont
  {H.}~\bibnamefont {{Estell{\'e}s}}}, \bibinfo {author} {\bibfnamefont
  {G.}~\bibnamefont {{Pratten}}}, \bibinfo {author} {\bibfnamefont
  {A.}~\bibnamefont {{Ramos-Buades}}}, \bibinfo {author} {\bibfnamefont
  {M.}~\bibnamefont {{Mateu-Lucena}}}, \ and\ \bibinfo {author} {\bibfnamefont
  {R.}~\bibnamefont {{Jaume}}},\ }\href {\doibase 10.1103/PhysRevD.102.064002}
  {\bibfield  {journal} {\bibinfo  {journal} {\prd}\ }\textbf {\bibinfo
  {volume} {102}},\ \bibinfo {eid} {064002} (\bibinfo {year} {2020})},\ \Eprint
  {http://arxiv.org/abs/2001.10914} {arXiv:2001.10914 [gr-qc]} \BibitemShut
  {NoStop}%
\bibitem [{\citenamefont {Welch}(1967)}]{1161901}%
  \BibitemOpen
  \bibfield  {author} {\bibinfo {author} {\bibfnamefont {P.}~\bibnamefont
  {Welch}},\ }\href {\doibase 10.1109/TAU.1967.1161901} {\bibfield  {journal}
  {\bibinfo  {journal} {IEEE Transactions on Audio and Electroacoustics}\
  }\textbf {\bibinfo {volume} {15}},\ \bibinfo {pages} {70} (\bibinfo {year}
  {1967})}\BibitemShut {NoStop}%
\bibitem [{\citenamefont {{Schutz}}(2011)}]{Sch11}%
  \BibitemOpen
  \bibfield  {author} {\bibinfo {author} {\bibfnamefont {B.~F.}\ \bibnamefont
  {{Schutz}}},\ }\href {\doibase 10.1088/0264-9381/28/12/125023} {\bibfield
  {journal} {\bibinfo  {journal} {Classical and Quantum Gravity}\ }\textbf
  {\bibinfo {volume} {28}},\ \bibinfo {eid} {125023} (\bibinfo {year}
  {2011})},\ \Eprint {http://arxiv.org/abs/1102.5421} {arXiv:1102.5421
  [astro-ph.IM]} \BibitemShut {NoStop}%
\bibitem [{\citenamefont {{Singh}}\ \emph {et~al.}(2023)\citenamefont
  {{Singh}}, \citenamefont {{Kapadia}}, \citenamefont {{Vijaykumar}},\ and\
  \citenamefont {{Ajith}}}]{Sin23_HM_Populations}%
  \BibitemOpen
  \bibfield  {author} {\bibinfo {author} {\bibfnamefont {M.~K.}\ \bibnamefont
  {{Singh}}}, \bibinfo {author} {\bibfnamefont {S.~J.}\ \bibnamefont
  {{Kapadia}}}, \bibinfo {author} {\bibfnamefont {A.}~\bibnamefont
  {{Vijaykumar}}}, \ and\ \bibinfo {author} {\bibfnamefont {P.}~\bibnamefont
  {{Ajith}}},\ }\href {\doibase 10.48550/arXiv.2312.07376} {\bibfield
  {journal} {\bibinfo  {journal} {arXiv e-prints}\ ,\ \bibinfo {eid}
  {arXiv:2312.07376}} (\bibinfo {year} {2023})},\ \Eprint
  {http://arxiv.org/abs/2312.07376} {arXiv:2312.07376 [gr-qc]} \BibitemShut
  {NoStop}%
\bibitem [{\citenamefont {{Ajith}}\ \emph {et~al.}(2014)\citenamefont
  {{Ajith}}, \citenamefont {{Fotopoulos}}, \citenamefont {{Privitera}},
  \citenamefont {{Neunzert}}, \citenamefont {{Mazumder}},\ and\ \citenamefont
  {{Weinstein}}}]{Aji14_TemplateBank}%
  \BibitemOpen
  \bibfield  {author} {\bibinfo {author} {\bibfnamefont {P.}~\bibnamefont
  {{Ajith}}}, \bibinfo {author} {\bibfnamefont {N.}~\bibnamefont
  {{Fotopoulos}}}, \bibinfo {author} {\bibfnamefont {S.}~\bibnamefont
  {{Privitera}}}, \bibinfo {author} {\bibfnamefont {A.}~\bibnamefont
  {{Neunzert}}}, \bibinfo {author} {\bibfnamefont {N.}~\bibnamefont
  {{Mazumder}}}, \ and\ \bibinfo {author} {\bibfnamefont {A.~J.}\ \bibnamefont
  {{Weinstein}}},\ }\href {\doibase 10.1103/PhysRevD.89.084041} {\bibfield
  {journal} {\bibinfo  {journal} {\prd}\ }\textbf {\bibinfo {volume} {89}},\
  \bibinfo {eid} {084041} (\bibinfo {year} {2014})},\ \Eprint
  {http://arxiv.org/abs/1210.6666} {arXiv:1210.6666 [gr-qc]} \BibitemShut
  {NoStop}%
\bibitem [{\citenamefont {{Brown}}\ \emph {et~al.}(2012)\citenamefont
  {{Brown}}, \citenamefont {{Harry}}, \citenamefont {{Lundgren}},\ and\
  \citenamefont {{Nitz}}}]{Brown12_GeometricPlacement}%
  \BibitemOpen
  \bibfield  {author} {\bibinfo {author} {\bibfnamefont {D.~A.}\ \bibnamefont
  {{Brown}}}, \bibinfo {author} {\bibfnamefont {I.}~\bibnamefont {{Harry}}},
  \bibinfo {author} {\bibfnamefont {A.}~\bibnamefont {{Lundgren}}}, \ and\
  \bibinfo {author} {\bibfnamefont {A.~H.}\ \bibnamefont {{Nitz}}},\ }\href
  {\doibase 10.1103/PhysRevD.86.084017} {\bibfield  {journal} {\bibinfo
  {journal} {\prd}\ }\textbf {\bibinfo {volume} {86}},\ \bibinfo {eid} {084017}
  (\bibinfo {year} {2012})},\ \Eprint {http://arxiv.org/abs/1207.6406}
  {arXiv:1207.6406 [gr-qc]} \BibitemShut {NoStop}%
\bibitem [{\citenamefont {Roulet}\ \emph {et~al.}(2019)\citenamefont {Roulet},
  \citenamefont {Dai}, \citenamefont {Venumadhav}, \citenamefont {Zackay},\
  and\ \citenamefont {Zaldarriaga}}]{ias_template_bank_PSD_roulet2019}%
  \BibitemOpen
  \bibfield  {author} {\bibinfo {author} {\bibfnamefont {J.}~\bibnamefont
  {Roulet}}, \bibinfo {author} {\bibfnamefont {L.}~\bibnamefont {Dai}},
  \bibinfo {author} {\bibfnamefont {T.}~\bibnamefont {Venumadhav}}, \bibinfo
  {author} {\bibfnamefont {B.}~\bibnamefont {Zackay}}, \ and\ \bibinfo {author}
  {\bibfnamefont {M.}~\bibnamefont {Zaldarriaga}},\ }\href {\doibase
  10.1103/PhysRevD.99.123022} {\bibfield  {journal} {\bibinfo  {journal} {Phys.
  Rev. D}\ }\textbf {\bibinfo {volume} {99}},\ \bibinfo {pages} {123022}
  (\bibinfo {year} {2019})}\BibitemShut {NoStop}%
\bibitem [{\citenamefont {{Hanna}}\ \emph {et~al.}(2023)\citenamefont {{Hanna}}
  \emph {et~al.}}]{Han23_GeometricPlacement}%
  \BibitemOpen
  \bibfield  {author} {\bibinfo {author} {\bibfnamefont {C.}~\bibnamefont
  {{Hanna}}} \emph {et~al.},\ }\href {\doibase 10.1103/PhysRevD.108.042003}
  {\bibfield  {journal} {\bibinfo  {journal} {\prd}\ }\textbf {\bibinfo
  {volume} {108}},\ \bibinfo {eid} {042003} (\bibinfo {year} {2023})},\ \Eprint
  {http://arxiv.org/abs/2209.11298} {arXiv:2209.11298 [gr-qc]} \BibitemShut
  {NoStop}%
\bibitem [{\citenamefont {Pedregosa}\ \emph {et~al.}(2011)\citenamefont
  {Pedregosa}, \citenamefont {Varoquaux}, \citenamefont {Gramfort},
  \citenamefont {Michel}, \citenamefont {Thirion}, \citenamefont {Grisel},
  \citenamefont {Blondel}, \citenamefont {Prettenhofer}, \citenamefont {Weiss},
  \citenamefont {Dubourg}, \citenamefont {Vanderplas}, \citenamefont {Passos},
  \citenamefont {Cournapeau}, \citenamefont {Brucher}, \citenamefont {Perrot},\
  and\ \citenamefont {{{\'E}}douard Duchesnay}}]{scikit_learn}%
  \BibitemOpen
  \bibfield  {author} {\bibinfo {author} {\bibfnamefont {F.}~\bibnamefont
  {Pedregosa}}, \bibinfo {author} {\bibfnamefont {G.}~\bibnamefont
  {Varoquaux}}, \bibinfo {author} {\bibfnamefont {A.}~\bibnamefont {Gramfort}},
  \bibinfo {author} {\bibfnamefont {V.}~\bibnamefont {Michel}}, \bibinfo
  {author} {\bibfnamefont {B.}~\bibnamefont {Thirion}}, \bibinfo {author}
  {\bibfnamefont {O.}~\bibnamefont {Grisel}}, \bibinfo {author} {\bibfnamefont
  {M.}~\bibnamefont {Blondel}}, \bibinfo {author} {\bibfnamefont
  {P.}~\bibnamefont {Prettenhofer}}, \bibinfo {author} {\bibfnamefont
  {R.}~\bibnamefont {Weiss}}, \bibinfo {author} {\bibfnamefont
  {V.}~\bibnamefont {Dubourg}}, \bibinfo {author} {\bibfnamefont
  {J.}~\bibnamefont {Vanderplas}}, \bibinfo {author} {\bibfnamefont
  {A.}~\bibnamefont {Passos}}, \bibinfo {author} {\bibfnamefont
  {D.}~\bibnamefont {Cournapeau}}, \bibinfo {author} {\bibfnamefont
  {M.}~\bibnamefont {Brucher}}, \bibinfo {author} {\bibfnamefont
  {M.}~\bibnamefont {Perrot}}, \ and\ \bibinfo {author} {\bibnamefont
  {{{\'E}}douard Duchesnay}},\ }\href
  {http://jmlr.org/papers/v12/pedregosa11a.html} {\bibfield  {journal}
  {\bibinfo  {journal} {Journal of Machine Learning Research}\ }\textbf
  {\bibinfo {volume} {12}},\ \bibinfo {pages} {2825} (\bibinfo {year}
  {2011})}\BibitemShut {NoStop}%
\bibitem [{\citenamefont {Owen}\ and\ \citenamefont
  {Sathyaprakash}(1999)}]{Owen:1998dk}%
  \BibitemOpen
  \bibfield  {author} {\bibinfo {author} {\bibfnamefont {B.~J.}\ \bibnamefont
  {Owen}}\ and\ \bibinfo {author} {\bibfnamefont {B.~S.}\ \bibnamefont
  {Sathyaprakash}},\ }\href {\doibase 10.1103/PhysRevD.60.022002} {\bibfield
  {journal} {\bibinfo  {journal} {Phys. Rev. D}\ }\textbf {\bibinfo {volume}
  {60}},\ \bibinfo {pages} {022002} (\bibinfo {year} {1999})},\ \Eprint
  {http://arxiv.org/abs/gr-qc/9808076} {arXiv:gr-qc/9808076} \BibitemShut
  {NoStop}%
\bibitem [{\citenamefont {{Philcox}}\ \emph {et~al.}(2021)\citenamefont
  {{Philcox}}, \citenamefont {{Ivanov}}, \citenamefont {{Zaldarriaga}},
  \citenamefont {{Simonovi{\'c}}},\ and\ \citenamefont
  {{Schmittfull}}}]{Phi21_PCA}%
  \BibitemOpen
  \bibfield  {author} {\bibinfo {author} {\bibfnamefont {O.~H.~E.}\
  \bibnamefont {{Philcox}}}, \bibinfo {author} {\bibfnamefont {M.~M.}\
  \bibnamefont {{Ivanov}}}, \bibinfo {author} {\bibfnamefont {M.}~\bibnamefont
  {{Zaldarriaga}}}, \bibinfo {author} {\bibfnamefont {M.}~\bibnamefont
  {{Simonovi{\'c}}}}, \ and\ \bibinfo {author} {\bibfnamefont {M.}~\bibnamefont
  {{Schmittfull}}},\ }\href {\doibase 10.1103/PhysRevD.103.043508} {\bibfield
  {journal} {\bibinfo  {journal} {\prd}\ }\textbf {\bibinfo {volume} {103}},\
  \bibinfo {eid} {043508} (\bibinfo {year} {2021})},\ \Eprint
  {http://arxiv.org/abs/2009.03311} {arXiv:2009.03311 [astro-ph.CO]}
  \BibitemShut {NoStop}%
\bibitem [{\citenamefont {{Cabanac}}\ \emph {et~al.}(2002)\citenamefont
  {{Cabanac}}, \citenamefont {{de Lapparent}},\ and\ \citenamefont
  {{Hickson}}}]{Cab02_PCA}%
  \BibitemOpen
  \bibfield  {author} {\bibinfo {author} {\bibfnamefont {R.~A.}\ \bibnamefont
  {{Cabanac}}}, \bibinfo {author} {\bibfnamefont {V.}~\bibnamefont {{de
  Lapparent}}}, \ and\ \bibinfo {author} {\bibfnamefont {P.}~\bibnamefont
  {{Hickson}}},\ }\href {\doibase 10.1051/0004-6361:20020665} {\bibfield
  {journal} {\bibinfo  {journal} {\aap}\ }\textbf {\bibinfo {volume} {389}},\
  \bibinfo {pages} {1090} (\bibinfo {year} {2002})},\ \Eprint
  {http://arxiv.org/abs/astro-ph/0206062} {arXiv:astro-ph/0206062 [astro-ph]}
  \BibitemShut {NoStop}%
\bibitem [{\citenamefont {{Medeiros}}\ \emph {et~al.}(2023)\citenamefont
  {{Medeiros}}, \citenamefont {{Psaltis}}, \citenamefont {{Lauer}},\ and\
  \citenamefont {{{\"O}zel}}}]{Lia23_EHT}%
  \BibitemOpen
  \bibfield  {author} {\bibinfo {author} {\bibfnamefont {L.}~\bibnamefont
  {{Medeiros}}}, \bibinfo {author} {\bibfnamefont {D.}~\bibnamefont
  {{Psaltis}}}, \bibinfo {author} {\bibfnamefont {T.~R.}\ \bibnamefont
  {{Lauer}}}, \ and\ \bibinfo {author} {\bibfnamefont {F.}~\bibnamefont
  {{{\"O}zel}}},\ }\href {\doibase 10.3847/2041-8213/acc32d} {\bibfield
  {journal} {\bibinfo  {journal} {\apjl}\ }\textbf {\bibinfo {volume} {947}},\
  \bibinfo {eid} {L7} (\bibinfo {year} {2023})},\ \Eprint
  {http://arxiv.org/abs/2304.06079} {arXiv:2304.06079 [astro-ph.HE]}
  \BibitemShut {NoStop}%
\bibitem [{\citenamefont {Venumadhav}\ \emph {et~al.}(2019)\citenamefont
  {Venumadhav}, \citenamefont {Zackay}, \citenamefont {Roulet}, \citenamefont
  {Dai},\ and\ \citenamefont
  {Zaldarriaga}}]{ias_pipeline_o1_catalog_new_search_prd2019}%
  \BibitemOpen
  \bibfield  {author} {\bibinfo {author} {\bibfnamefont {T.}~\bibnamefont
  {Venumadhav}}, \bibinfo {author} {\bibfnamefont {B.}~\bibnamefont {Zackay}},
  \bibinfo {author} {\bibfnamefont {J.}~\bibnamefont {Roulet}}, \bibinfo
  {author} {\bibfnamefont {L.}~\bibnamefont {Dai}}, \ and\ \bibinfo {author}
  {\bibfnamefont {M.}~\bibnamefont {Zaldarriaga}},\ }\href {\doibase
  10.1103/PhysRevD.100.023011} {\bibfield  {journal} {\bibinfo  {journal}
  {Phys. Rev.}\ }\textbf {\bibinfo {volume} {D100}},\ \bibinfo {pages} {023011}
  (\bibinfo {year} {2019})},\ \Eprint {http://arxiv.org/abs/1902.10341}
  {arXiv:1902.10341 [astro-ph.IM]} \BibitemShut {NoStop}%
\bibitem [{\citenamefont {{Cheung}}\ \emph {et~al.}()\citenamefont {{Cheung}},
  \citenamefont {{Wadekar}} \emph {et~al.}}]{Che24_IMRI_search}%
  \BibitemOpen
  \bibfield  {author} {\bibinfo {author} {\bibfnamefont {M.~H.-Y.}\
  \bibnamefont {{Cheung}}}, \bibinfo {author} {\bibfnamefont {D.}~\bibnamefont
  {{Wadekar}}},  \emph {et~al.},\ }\href@noop {} {\bibinfo  {journal} {in
  preparation}\ }\BibitemShut {NoStop}%
\bibitem [{\citenamefont {{Mishra}}\ \emph {et~al.}(2016)\citenamefont
  {{Mishra}}, \citenamefont {{Kela}}, \citenamefont {{Arun}},\ and\
  \citenamefont {{Faye}}}]{Mis16}%
  \BibitemOpen
\bibfield  {journal} {  }\bibfield  {author} {\bibinfo {author} {\bibfnamefont
  {C.~K.}\ \bibnamefont {{Mishra}}}, \bibinfo {author} {\bibfnamefont
  {A.}~\bibnamefont {{Kela}}}, \bibinfo {author} {\bibfnamefont {K.~G.}\
  \bibnamefont {{Arun}}}, \ and\ \bibinfo {author} {\bibfnamefont
  {G.}~\bibnamefont {{Faye}}},\ }\href {\doibase 10.1103/PhysRevD.93.084054}
  {\bibfield  {journal} {\bibinfo  {journal} {\prd}\ }\textbf {\bibinfo
  {volume} {93}},\ \bibinfo {eid} {084054} (\bibinfo {year} {2016})},\ \Eprint
  {http://arxiv.org/abs/1601.05588} {arXiv:1601.05588 [gr-qc]} \BibitemShut
  {NoStop}%
\bibitem [{\citenamefont {{Borhanian}}\ \emph {et~al.}(2020)\citenamefont
  {{Borhanian}}, \citenamefont {{Arun}}, \citenamefont {{Pfeiffer}},\ and\
  \citenamefont {{Sathyaprakash}}}]{Boh20_HM_PN}%
  \BibitemOpen
  \bibfield  {author} {\bibinfo {author} {\bibfnamefont {S.}~\bibnamefont
  {{Borhanian}}}, \bibinfo {author} {\bibfnamefont {K.~G.}\ \bibnamefont
  {{Arun}}}, \bibinfo {author} {\bibfnamefont {H.~P.}\ \bibnamefont
  {{Pfeiffer}}}, \ and\ \bibinfo {author} {\bibfnamefont {B.~S.}\ \bibnamefont
  {{Sathyaprakash}}},\ }\href {\doibase 10.1088/1361-6382/ab6a21} {\bibfield
  {journal} {\bibinfo  {journal} {Classical and Quantum Gravity}\ }\textbf
  {\bibinfo {volume} {37}},\ \bibinfo {eid} {065006} (\bibinfo {year}
  {2020})},\ \Eprint {http://arxiv.org/abs/1901.08516} {arXiv:1901.08516
  [gr-qc]} \BibitemShut {NoStop}%
\bibitem [{\citenamefont {Neyman}\ and\ \citenamefont
  {Pearson}(1933)}]{neymanpearson}%
  \BibitemOpen
  \bibfield  {author} {\bibinfo {author} {\bibfnamefont {J.}~\bibnamefont
  {Neyman}}\ and\ \bibinfo {author} {\bibfnamefont {E.~S.}\ \bibnamefont
  {Pearson}},\ }\href@noop {} {\bibfield  {journal} {\bibinfo  {journal}
  {Philosophical Transactions of the Royal Society of London. Series A,
  Containing Papers of a Mathematical or Physical Character}\ }\textbf
  {\bibinfo {volume} {231}},\ \bibinfo {pages} {289} (\bibinfo {year}
  {1933})}\BibitemShut {NoStop}%
\bibitem [{\citenamefont {{McIsaac}}\ \emph {et~al.}(2023)\citenamefont
  {{McIsaac}}, \citenamefont {{Hoy}},\ and\ \citenamefont {{Harry}}}]{McI23}%
  \BibitemOpen
  \bibfield  {author} {\bibinfo {author} {\bibfnamefont {C.}~\bibnamefont
  {{McIsaac}}}, \bibinfo {author} {\bibfnamefont {C.}~\bibnamefont {{Hoy}}}, \
  and\ \bibinfo {author} {\bibfnamefont {I.}~\bibnamefont {{Harry}}},\ }\href
  {\doibase 10.48550/arXiv.2303.17364} {\bibfield  {journal} {\bibinfo
  {journal} {arXiv e-prints}\ ,\ \bibinfo {eid} {arXiv:2303.17364}} (\bibinfo
  {year} {2023})},\ \Eprint {http://arxiv.org/abs/2303.17364} {arXiv:2303.17364
  [gr-qc]} \BibitemShut {NoStop}%
\end{thebibliography}%

\end{document}